\documentclass[a4paper,11pt]{article}
\pdfoutput=1 

\usepackage{jcappub}

\usepackage[T1]{fontenc} 
\usepackage{amsmath,amssymb}
\usepackage{dsfont}
\usepackage{mathtools}
\usepackage{animate,media9}
\usepackage{slashed}
\usepackage{braket}
\usepackage{subfigure}
\usepackage{slashed} 
\usepackage{hyperref}

\usepackage{graphicx}
\usepackage[usenames,dvipsnames]{pstricks}
\usepackage{subfigure}
\usepackage{epsfig}
\usepackage{graphicx}
\usepackage{xcolor,color}
\usepackage{geometry}
\usepackage{bm}
\usepackage[section]{placeins}

\newcommand{\Moller}{M{\o}ller }

\title{Gravothermal evolution of dark matter halos with differential elastic scattering }

\author[a]{Daneng Yang,}
\author[a]{Hai-Bo Yu}

\emailAdd{danengy@ucr.edu}
\emailAdd{haiboyu@ucr.edu}

\affiliation[a]{Department of Physics and Astronomy, University of California, Riverside, California 92521, USA}

\abstract{
We study gravothermal evolution of dark matter halos in the presence of differential self-scattering that has strong velocity and angular dependencies. We design controlled N-body simulations to model Rutherford and \Moller scatterings in the halo, and follow its evolution in both core-expansion and -collapse phases. The simulations show the commonly-used transfer cross section underestimates the effects of dark matter self-interactions, but the viscosity cross section provides an accurate approximation for modeling angular-dependent dark matter scattering. We investigate thermodynamic properties of the halo, and find that the three moments of the Boltzmann equation under the fluid approximation are satisfied. We further propose a constant effective cross section, which integrates over the halo's characteristic velocity dispersion with weighting kernels motivated by kinetic theory of heat conduction. The effective cross section provides a good approximation to differential self-scattering for most of the halo evolution. It indicates that we can map astrophysical constraints on a constant self-interacting cross section to an SIDM model with velocity- and angular-dependent scatterings. 

}

\begin{document}
\maketitle

\flushbottom

\tableofcontents

\section{Introduction}

Self-interacting dark matter (SIDM) is a well-motivated scenario where dark matter is assumed to have strong self-interactions, analogous to the nuclear interactions, see~\cite{Tulin:2017ara} for a review. In SIDM, the interactions can thermalize the inner regions of dark matter halos, and change its structure accordingly~\cite{Spergel:1999mh,Dave:2000ar,Vogelsberger:2012ku,Rocha:2012jg}, while it keeps all the success of the prevailing cold dark matter model on large scales. Recent studies show that SIDM predicts diverse dark matter distributions in both main~\cite{Kaplinghat:2013xca,Kamada:2016euw,Creasey:2016jaq,Robertson:2017mgj,Ren:2018jpt,Sameie:2021ang} and satellite~\cite{Nishikawa:2019lsc,Sameie:2019zfo,Kahlhoefer:2019oyt,Yang:2020iya,Yang:2021kdf} halos, a feature that is favored in explaining observations on galactic scales, see~\cite{Oman:2015xda,Bullock:2017xww,Salucci:2018hqu,Kaplinghat:2019svz}. 

Most SIDM studies assume a constant self-interacting cross section. However, recent work shows that the required cross section per mass is $\sigma/m_\chi\gtrsim1\rm ~cm^2/g$ to explain the observations of galaxies, see~\cite{Tulin:2017ara}, while it is $\sigma/m_\chi\lesssim0.1\rm ~cm^2/g$ in galaxy clusters~\cite{Kaplinghat:2015aga,Harvey:2015hha,Sagunski:2020spe,Andrade:2020lqq}. Reconciling these observations requires the cross section to be velocity-dependent. In addition, a strong velocity-dependent cross section is needed for explaining dark matter densities of dwarf spheroidal galaxies of the Milky Way in SIDM~\cite{Valli:2017ktb,Zavala:2019sjk,Correa:2020qam,Ebisu:2021bjh,Jiang:2021foz,Silverman:2022bhs}. From the perspective of particle physics, it is almost inevitable to consider differential scattering cross sections which depend on both relative velocity and scattering angle~\cite{Feng:2009mn,Feng:2009hw,Buckley:2009in,Loeb:2010gj,Tulin:2013teo,Cyr-Racine:2015ihg,Agrawal:2016quu}. Recent simulations have implemented velocity-dependent cross sections~\cite{Vogelsberger:2012ku,Vogelsberger:2015gpr,Robertson:2018anx,Nadler:2020ulu,Turner:2020vlf,Bhattacharyya:2021vyd,Zeng:2021ldo}. In particular, Ref.~\cite{Robertson:2016qef} investigated how well an isotropic cross section could capture the evolution of an SIDM halo if the actual scattering is anisotropic. Refs.~\cite{Kahlhoefer:2013dca,Fischer:2020uxh, Fischer:2022rko} simulated frequent self-interactions with small angles. 

In this work, we perform high-resolution controlled N-body simulations to study gravothermal evolution of dark matter halos in the presence of differential self-scattering, which has strong velocity and angular dependencies. We consider Rutherford and \Moller scatterings in the halo, and follow the evolution of its density and velocity dispersion in both core-expansion and -collapse phases. We also perform simulations with the transfer and viscosity cross section, which are velocity-dependent, but angular-independent. We will show that the viscosity cross section provides a good approximation for modeling differential self-interactions for both Rutherford and \Moller scatterings. This result holds in the expansion and collapse phases. 

We study thermodynamic properties of the simulated halo and understand its evolution history from the perspective of thermodynamics. In particular, we reconstruct radial profiles of the luminosity, specific heat, entropy change rate and heat conductivity, and centripetal acceleration. We will show that the three moments of the Boltzmann equation under the fluid approximation are satisfied for the simulated halo, and heat conduction is in the long-mean-free-path regime. We further propose an effective cross section for modeling halo evolution, with weighting kernels as $\sin^2\theta$ and $v^5$, motivated by kinetic theory of heat conduction. For a given halo, we specify a single characteristic velocity, such that the effective cross section can be expressed using a constant value, which does not explicitly depend on the velocity and angle. We will use simulations to confirm the validity of the constant effective cross section. 

The rest of the paper is organized as follows. In Sec.~\ref{sec:xs}, we introduce the microscopic description of dark matter self-interactions, discuss our simulation setup and show numerical comparisons among the simulation results. In Sec.~\ref{sec:thermal}, we study thermodynamic properties of the halo. In Sec.~\ref{sec:sigmaeff}, we introduce a constant effective cross section and test it with the simulations. In Sec.~\ref{sec:con}, we conclude. In Appendix~\ref{app:sidm}, we provide the comparison that validates our SIDM module implemented in N-body simulations. In Appendix~\ref{app:conv}, we discuss convergence tests of our SIDM simulations.

\section{Differential dark matter self-scattering}
\label{sec:xs}

In this section, we will discuss Rutherford and M{\o}ller scatterings for SIDM and provide essential formulae for calculating various self-scattering cross sections. In addition, we will discuss implementations of N-body simulations for modeling velocity- and angular-dependent dark matter self-interactions and show numerical comparisons. 

\subsection{Differential scattering cross sections}

We consider a scenario where a light gauge boson mediates elastic dark matter self-interactions, see~\cite{Tulin:2017ara} for a review of SIDM models. Depending on the production mechanism, dark matter could be symmetric or asymmetric. In the symmetric case, both particles ($\chi$) and anti-particles ($\bar{\chi}$) present in the halo, there are three types of scattering processes, i.e., $\chi\chi\rightarrow\chi\chi$, $\bar{\chi}\bar{\chi}\rightarrow\bar{\chi}\bar{\chi}$ and $\chi\bar{\chi}\rightarrow\chi\bar{\chi}$. In the asymmetric case where the halo is dominated by one species, say $\chi$, the relevant process is $\chi\chi\rightarrow\chi\chi$. If two initial states are different, we only need to include a $t$-channel Feynman diagram in calculating the scattering amplitude at the leading order, analogous to Rutherford scattering in nuclear physics. However, if the initial states are indistinguishable, both $t$- and $u$-channel diagrams contribute, similar to M{\o}ller scattering. Thus in general both Rutherford and M{\o}ller scatterings are relevant for symmetric SIDM, while the latter is relevant for asymmetric SIDM.

Consider an SIDM particle ($\chi$) that couples to a light gauge boson ($\phi$) with an interaction strength of $g_\chi$ as $ig\bar{\chi}\gamma^\mu\chi\phi_\mu$. In the weakly-coupled perturbative limit, the differential cross section in the center of momentum frame for Rutherford scattering ($\chi\bar{\chi}\rightarrow\chi\bar{\chi}$) is~\cite{Feng:2009hw,Ibe:2009mk} 
\begin{eqnarray}
\label{eq:xsr}
\frac{d\sigma}{d \cos\theta} = \frac{\sigma_{0}w^4}{2\left[w^2+{v^{2}}\sin^2(\theta/2)\right]^2 },
\end{eqnarray}
where $\sigma_0\equiv g^4_\chi/(4\pi m^2_\chi w^4)=4\pi\alpha^2_\chi/(m^2_\chi w^4)$ with $\alpha_\chi\equiv g^2_\chi/4\pi$, $w\equiv {m_\phi}c/{m_\chi}$, $v$ is the relative velocity between the two initial particles, and $\theta$ the scattering angle. We have assumed $m_\chi$ and $m_\phi$ to be masses of dark matter and mediator particles, respectively. In addition, in writing the expression in Eq.~\ref{eq:xsr}, we have used the parametrization with $\sigma_0$ and $w$ proposed in~\cite{Robertson:2016qef}. Integrating out the angular distribution, one obtains the total cross section as $\sigma_{\rm tot}= {\sigma_0}/(1+{v^2}/{w^2})$.

For \Moller scattering ($\chi\chi\rightarrow\chi\chi$), we calculate the differential cross section as
\begin{equation}
\label{eq:xsm}
\dfrac{d\sigma}{d \cos\theta} = \frac{\sigma_0 w^4 \left[\left(3 \cos^2\theta+1\right) v^4+4 v^2 w^2+4  w^4\right]}{\left(\sin^2\theta v^4+4 v^2 w^2+4 w^4\right)^2}, 
\end{equation}
where we have already included a symmetry factor of $1/2$ to take into account the fact that the particles in the final state are identical, and the scattering angle $\theta$ takes the value from $0$ to $\pi$ as in the case of Rutherford scattering; see also~\cite{Girmohanta:2022dog} for a derivation. The corresponding total scattering cross section is 
\begin{equation}
\sigma_{\rm tot}= \sigma_0 w^4 \left[\frac{1}{v^2 w^2+w^4}+\frac{1}{v^4+2   v^2 w^2} \ln \left(\frac{w^2}{v^2+w^2}\right)\right].
\end{equation}

\begin{figure*}
  \centering
  \includegraphics[height=4.5cm]{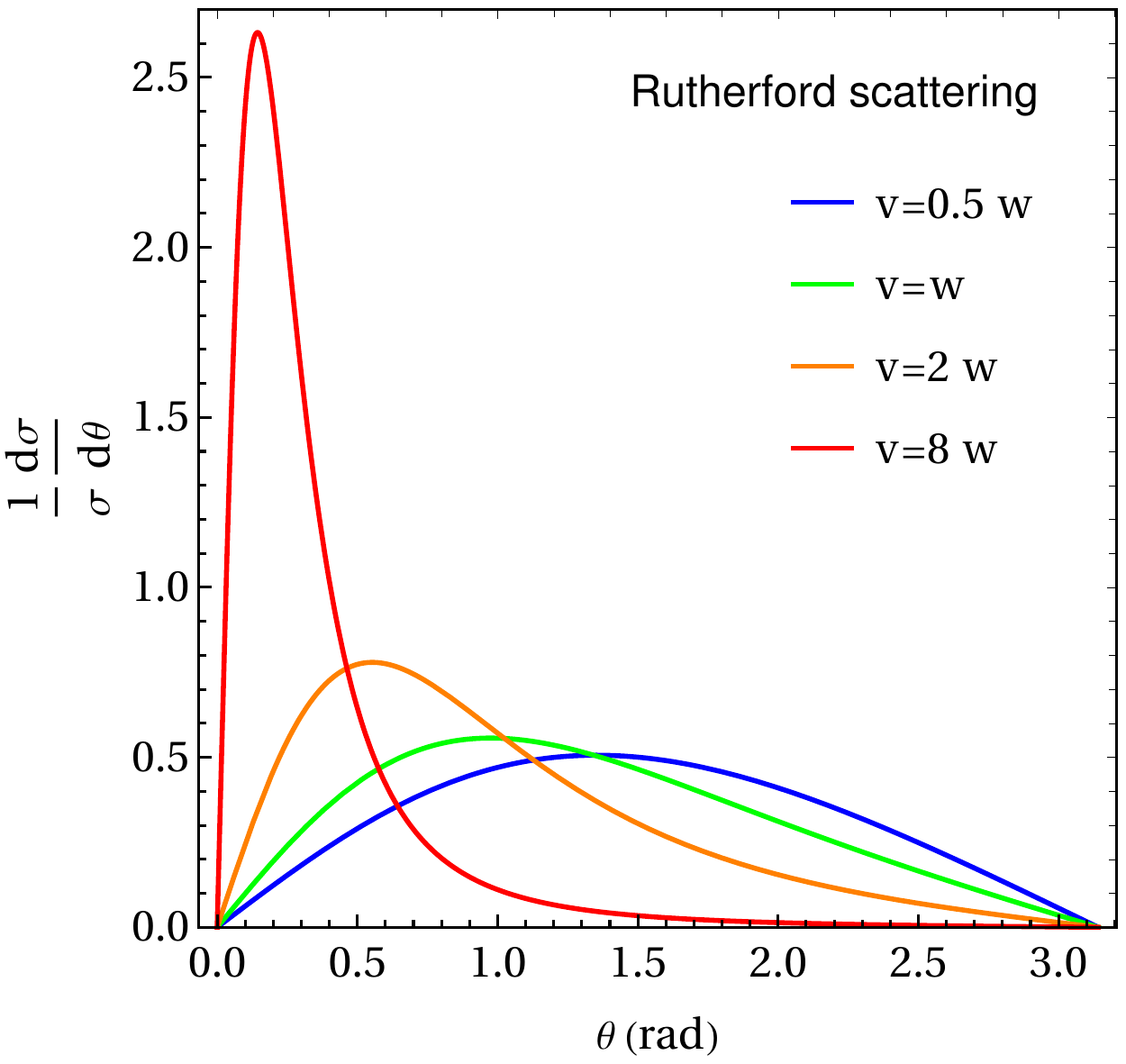}
  \includegraphics[height=4.5cm]{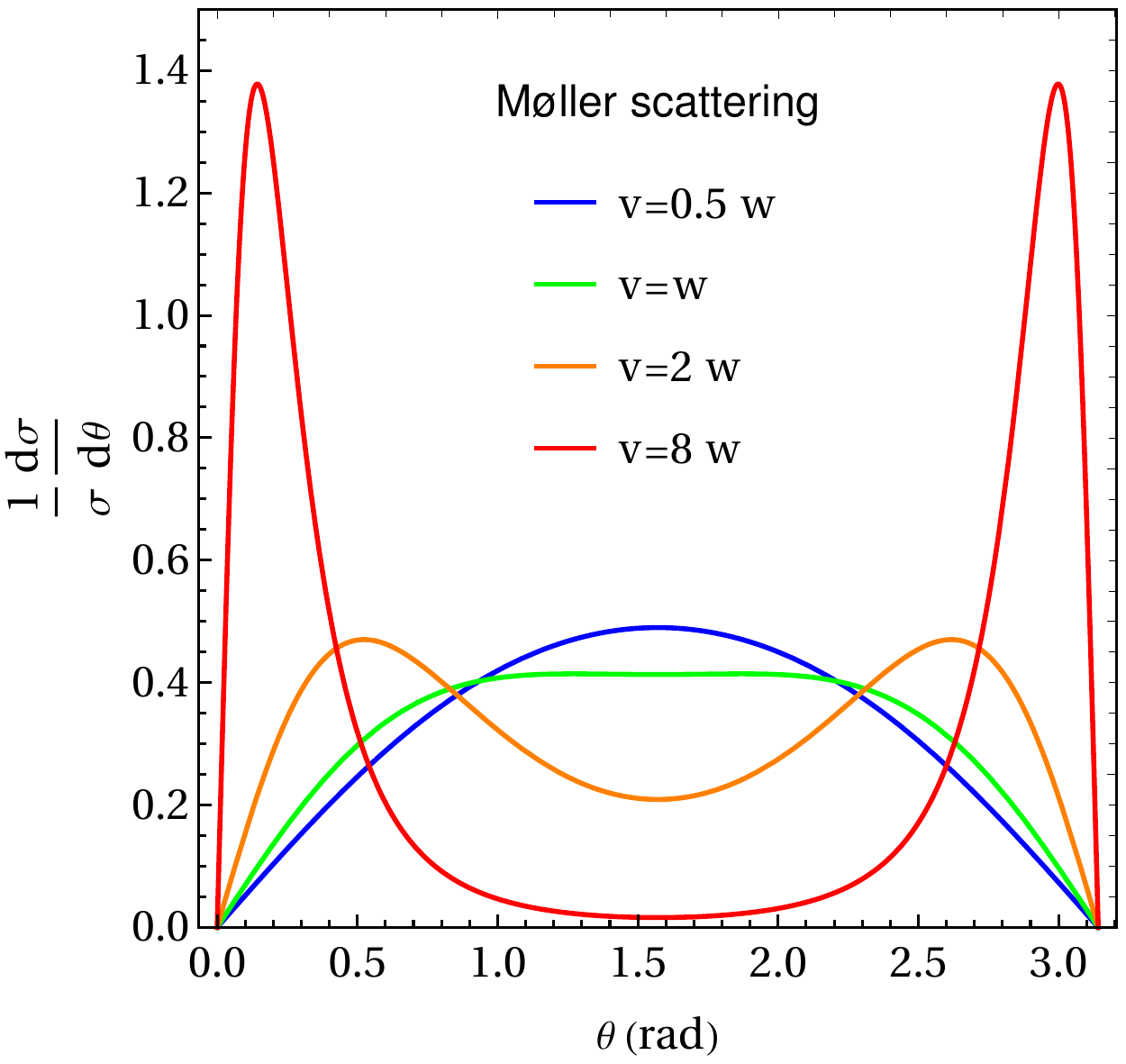}
    \includegraphics[height=4.5cm]{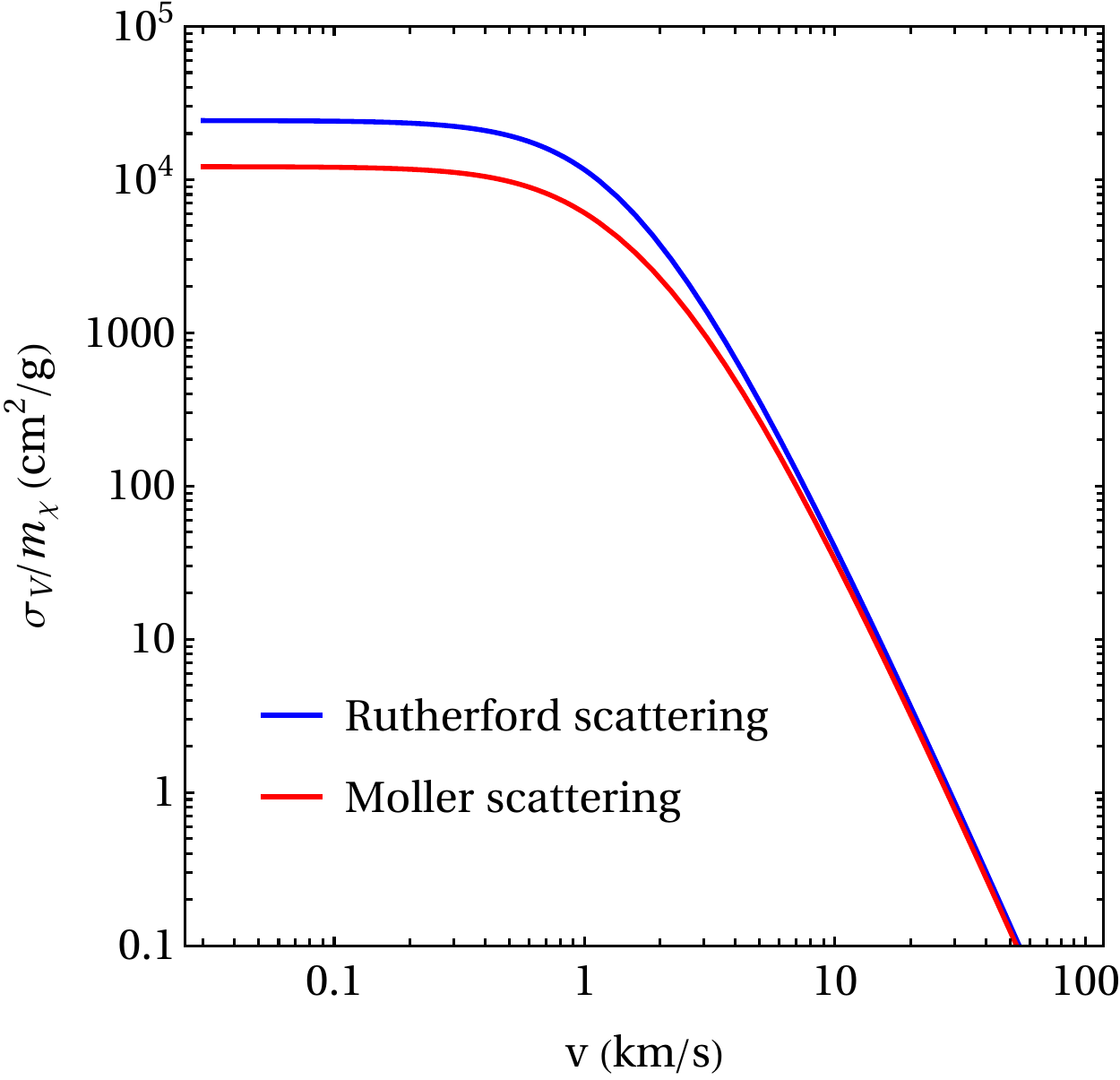}
  \caption{{\it Left:} Angular dependence of differential cross sections for Rutherford scattering {\it Middle:} Angular dependence of differential cross sections for \Moller scattering. In both panels, $v$ is the relative velocity of dark matter particles and $w=m_\phi c/m_\chi$. {\it Right:} Velocity dependence of viscosity cross sections for Rutherford and \Moller scatterings, where $\sigma_0/m_\chi=2.4\times10^4~{\rm cm^2/g}$ and $w=m_\phi c/m_\chi=1~{\rm km/s}$.  }
\label{fig:xscomp}
\end{figure*}

Fig.~\ref{fig:xscomp} shows the angular dependence of normalized differential cross sections for Rutherford (left panel) and \Moller scatterings (middle panel). In both cases, the dependence becomes significant as the velocity increases with respect to $w=m_\phi c/m_\chi$. For Rutherford scattering, the cross section peaks towards small angles ($\theta\rightarrow0$) and the significance increases with the relative velocity. For \Moller scattering, it peaks towards both small and large angles ($\theta\rightarrow0,~\pi$), and the distribution is symmetric around $\theta=\pi/2$. In addition, the interference effect manifests in the perpendicular direction $\theta=\pi/2$ when the velocity is low. 

It is well known that for Rutherford scattering the enhancement of the differential cross section in the forward direction ($\theta\rightarrow0$) is spurious in changing halo structure. Thus using the total cross section $\sigma_{\rm tot}$ is not a good measure as it likely overestimates the self-scattering effects. To regulate the forward scattering, a ``transfer'' cross section is often considered~\cite{Mohapatra:2001sx,Feng:2009hw,Buckley:2009in}
\begin{eqnarray}
\label{eq:transport}
\sigma_T = \int d \cos\theta (1-\cos\theta)\frac{d\sigma}{d\cos\theta}, 
\end{eqnarray}
where the factor $(1-\cos\theta)$ is related to the momentum transfer $\Delta p=-mv(1-\cos\theta)$ during the collision.\footnote{In kinetic theory of gases~\cite{lifshitz1995physical}, $\sigma_T$ in Eq.~\ref{eq:transport} is called the ``transport'' cross section that characterizes diffusion of colliding gaseous particles. In this paper, we refer to $\sigma_T$ as the ``transfer'' cross section.} The transfer cross section regulates forward collisions ($\theta\rightarrow0$), but it rewards backward ones ($\theta\rightarrow\pi$), which hardly change the halo structure. Another disadvantage is that the transfer cross section is not well defined when the interference between $t$- and $u$-channels is present as in \Moller scattering. Ref.~\cite{Tulin:2013teo} first suggested using a ``viscosity'' cross section to model dark matter self-interactions, see also~\cite{Cline:2013pca,Boddy:2016bbu,Blennow:2016gde,Alvarez:2019nwt,Colquhoun:2020adl},
\begin{equation}
\label{eq:xsrD}
\sigma_V = \frac{3}{2} \int d\cos\theta \sin^2\theta \frac{d\sigma}{d\cos\theta},
\end{equation}
where we have included a normalization factor of $3/2$ such that the relation $\sigma_{V}=\sigma_{\rm tot}$ holds for isotropic scattering. 
The ``viscosity'' cross section regulates both backward and forward scatterings. In addition, it weighs most the perpendicular direction $\theta=\pi/2$, at which the collisions thermalize the system mostly, as we will discuss later.  

For Rutherford scattering, the transfer and viscosity cross sections are
\begin{eqnarray}
\label{eq:xsrT}
\sigma_{T} &=& \frac{2 \sigma_0 w^4}{v^4} \left[\ln\left(1+\frac{v^2}{w^2}\right)-\frac{v^2}{v^2+w^2} \right], \\ 
\label{eq:xsrV}
\sigma_{V} &=& \frac{6 \sigma_0 w^6}{v^6}    \left[\left(2+\frac{v^2}{w^2}\right) \ln\left(1+\frac{v^2}{w^2}\right)-\frac{2 v^2}{w^2}\right],
\end{eqnarray}
respectively. For \Moller scattering, we calculate the corresponding viscosity cross section as
\begin{eqnarray}
\label{eq:xsm}
\sigma_{V} = \frac{3 \sigma_0 w^8}{v^8+2 v^6 w^2} \left[2 \left(5+5 \frac{v^2 w^2}{w^4} +\frac{v^4}{w^4} \right) \ln \left(1+\frac{v^2}{w^2}\right)-5 \left(\frac{v^4}{w^4}+2 \frac{v^2}{w^2}\right)\right].
\end{eqnarray}

Ref.~\cite{Robertson:2016qef} performed N-body simulations for Rutherford scattering to test whether the transfer cross section in Eq.~\ref{eq:xsrT} accurately describes the evolution of an SIDM halo for $t$-channel {\it anisotropic} scattering. They showed that the core size of a halo simulated using $\sigma_{T}$ is $20\%$ smaller, compared to the one using a differential cross section. The agreement can be reached within $5\%$ if one uses a modified version of the transfer cross section defined as $\sigma'_{T}=2\int^\pi_0(1-|\cos\theta|)(d\sigma/d\Omega)d\Omega$ proposed in~\cite{Kahlhoefer:2013dca}. 

In this work, we simulate gravothermal evolution of SIDM halos using the velocity- and angular-dependent differential cross section $d\sigma/d\cos\theta$, as well as angular-independent viscosity $\sigma_V$ and transfer $\sigma_T$ cross sections. We will show that $\sigma_V$ provides an excellent approximation for modeling differential dark matter self-interactions in both core-expansion and -collapse phases for Rutherford {\it and} \Moller scatterings. 

We choose the following model parameters in our simulations based on $d\sigma/d\cos\theta$ $\sigma_V$ and $\sigma_T$, $\sigma_0/m_\chi=2.4\times10^{4}~{\rm cm^2/g}$ and $w=1~{\rm km/s}$. With these parameters, we have $\sigma_{V}/m_\chi = 10~\rm cm^2/g$ for Rutherford scattering in a halo with $v=15~{\rm km/s}$. With such a large value of $\sigma_0/m_\chi$, we need to make sure that the perturbative approximation is valid. For a Yukawa interaction, the condition is $\alpha_\chi m_\chi/m_\phi<1$~\cite{Tulin:2013teo}. For example, we can consider $\alpha_\chi=10^{-6}$, $m_\chi=9.7~{\rm GeV}$ and $m_\phi=32~{\rm keV}$. With the choice of the model parameters, the self-scattering cross section has a strong velocity dependence and it is enhanced towards low velocities, while being consistent with constraints from cluster scales. Our simulations focus on halos with $v\sim10~{\rm km/s}$, as velocity-dependent SIDM is particularly interesting for dwarf halos.

Fig.~\ref{fig:xscomp} (right panel) shows the viscosity cross section vs. velocity for Rutherford (blue) and \Moller (red) scatterings. In the very low velocity limit $v\ll w$, $\sigma_V\approx\sigma_{\rm tot} \rightarrow\sigma_0$ and $\sigma_0/2$ for Rutherford and \Moller scatterings, respectively. For the latter case, there is a destructive interference and the cross section is reduced as the de-Broglie wavelength of dark matter particles ($1/m_\chi v$) is larger than the Yukawa interaction range ($1/m_\phi c$). In the opposite limit $v\gg w$, i.e., $m_\chi v\gg m_\phi c$, both cases have the same $\sigma_V$ that scales as $\sigma_V\propto 1/v^4$. This is the classical regime~\cite{Feng:2009hw} and the interference effect vanishes.

\subsection{Simulation setup}

We use controlled N-body simulations to test the accuracy of {\it isotropic} scattering cross sections $\sigma_V$ and $\sigma_T$ in capturing gravothermal evolution of isolated SIDM halos that involve {\it anisotropic} collisions. We develop an SIDM module and implement it to the public \texttt{GADGET-2} program~\cite{Springel:2005mi,Springel:2000yr}, following the instructions in~\cite{2017MNRAS.465...76M}. Our module uses similar techniques as in~\cite{Rocha:2012jg,Robertson:2016xjh} to model dark matter self-interactions. For each particle, we search for its neighbors within a sphere of radius that equals the gravitational softening length $h$ from its Cartesian coordinate positions. We select some of the neighbors to interact with the particle based on the scattering probability
\begin{equation}
\label{eq:probability}
{\cal P}_{ij} = \frac{1}{2 S_{ij}} \sigma(v_{ij}) v_{ij} W(r_{ij},h) \Delta t,  
\end{equation}
where $\sigma(v_{ij})$ is the total, transfer and viscosity cross sections for $d\sigma/d\cos\theta$, $\sigma_T$ and $\sigma_V$ simulations, respectively, $\Delta t$ a small time interval, $W(r,h)$ a weighting kernel, and the factor $1/2$ removes double counting from looping over the particle indices $i,j$. The factor $S_{i,j}$ equals two for identical $i,j$ particles and equals one otherwise. It removes double counting in the phase space of two identical particles. 

We choose the kernel function be the same as the smoothing kernel in~\texttt{GADGET-2}~\cite{Springel:2005mi}, which reads
\begin{eqnarray}
\label{eq:kernel}
W(r,h) = \frac{8}{\pi h^3} \left\{ 
\begin{array}{ll} 
1-6\left(\frac{r}{h} \right)^2 + 6 \left(\frac{r}{h} \right)^3, & 0 \leq \frac{r}{h} \leq \frac{1}{2}, \\
2 \left( 1-\frac{r}{h}\right)^3, & \frac{1}{2} < \frac{r}{h} \leq 1, \\ 
0, & \frac{r}{h} > 1. 
\end{array}
\right. \\  \nonumber
\end{eqnarray}
It is normalized such that $4\pi \int_0^h d x x^2 W(x,h) = 1$. We set the size of the kernel to be $h=2.8\epsilon$, where $\epsilon$ is the gravitational softening length, motivated by the test performed in~\cite{Robertson:2016xjh}. Note our choice of the kernel function is different from that in~\cite{Robertson:2016xjh}, where a top hat kernel was used, but consistent with~\cite{Rocha:2012jg}. We set the timestep $\Delta t$ in calculating the scattering probability in Eq.~\ref{eq:probability} to be the same as the one for calculating gravity in~\texttt{GADGET-2}. Our implementation allows a candidate particle to interact with multiple neighbors within one timestep $\Delta t$, and we randomize the ordering of the selected particles to scatter with the candidate. After the collision, we update kinematics of the colliding particles and ensure that both momentum and energy are conserved.

After determining the neighboring particles to interact, we need to model the angular distribution in dark matter collisions. For the $\sigma_V$ and $\sigma_T$ simulations, the scattering is isotropic and we use the standard method, see, e.g.,~\cite{Rocha:2012jg}. For \Moller scattering, we sample the angular distribution using the method of rejection sampling, which can be applied to any generic angular distribution. For Rutherford scattering, since the cumulative distribution function can be inverted analytically, we follow the method in~\cite{Robertson:2016xjh}, which is more efficient. We have verified that the two approaches lead to identical results in the Rutherford case.

In practice, we prepare the same initial state for both Rutherford and \Moller scatterings. Since simulation particles are distinguishable and collisions could occur among all of them, we do not directly include the symmetry factor $S_{ij}$ in Eq.~\ref{eq:probability} in calculating the scattering probability. Instead, we take the following procedure to interpret our simulation results for different scattering types. For the scenario with Rutherford scattering, a dark matter halo contains two distinct species and each of them comprises half of the simulation particles. Thus the actual cross section between the two species is four times larger than the simulated value ($\sigma_0$). For \Moller scattering, a halo contains one species and the actual cross section is a factor two larger than the simulated value ($\sigma_0$), in order to restore the symmetry factor $1/2$ for identical particles in the initial state when computing the scattering probability.

\begin{table}
\begin{center}
\begin{tabular}{c|c|c|c|c}
\hline
\hline
Name & $M_{200}$ ($\rm M_{\odot}$) & $c_{200}(z=2)$  &  $\rho_s \left(\frac{\rm M_{\odot}}{\rm kpc^3}\right)$  &  $r_s$ (kpc)   \\
\hline
BM1        & $1\times 10^{7}$            & 20.4                            &  $2.99\times 10^{8}$      &   0.108   \\
BM2        & $2\times 10^{7}$            & 19.7                         &  $2.74\times 10^{8}$      &   0.141     \\
BM3        & $3\times 10^{7}$            & 19.3                   &  $2.60\times 10^{8}$      &   0.164    \\
\hline
\hline
\end{tabular}
\caption{Parameters of the simulated halos. From the left to right columns: labeling name, halo mass, halo concentration, scale density, and scale radius. For all the initial halos, their concentration is four times the standard deviation higher than the cosmological median.
\label{tab:simbm} }
\end{center}
\end{table}
We consider three benchmark sets of initial halo parameters named as BM1, BM2, and BM3, see Table~\ref{tab:simbm}. 
To explore the full stage of gravothermal evolution, we choose the halos with a high concentration, based on the concentration-mass relation from cosmological simulations at redshift $z=2$~\cite{Dutton:2014xda}, such that the timescale for the onset of gravothermal collapse is considerably short. We further assume an initial Navarre-Frenk-White (NFW) density profile~\cite{1997ApJ...490..493N} and use the~\texttt {SpherIC} code~\cite{GarrisonKimmel:2013aq} to generate initial conditions. We set the softening length as $\epsilon = 4 r_{200}/\sqrt{N}$~\cite{Power:2002sw}, where $r_{200}=c_{200}r_s$ is the radius at which the average density is $200$ times the critical density of the universe, and $N$ is the number of simulation particles. The gravitational force between pairs of particles is Newtonian when their separation is larger than $2.8\epsilon$. As discussed, we set the size of the kernel function~\ref{eq:kernel} to be $h=2.8\epsilon$.

We have tested and validated our SIDM implementation using the results in~\cite{Robertson:2016qef}, see Appendix~\ref{app:sidm}. In addition, we have performed convergence tests for the model parameters we consider in this work, see Appendix~\ref{app:conv} for details. In particular, we have explored the number of simulation particles and the timestep that are required to achieve convergence. With our simulation setup, we find if the particle number $N$ and the accuracy parameter $\eta$ controlling timestep satisfy ($N=4\times10^6$, $\eta=0.025$) or ($N=10^6$, $\eta=0.0025$), the evolution of the halo central density converges. We will present our main simulation results, which pass the tests, based on the BM2 initial halo. Only in Sec.~\ref{sec:differenthalo}, we will compare simulation results with all three benchmark halos. When we show the evolution of the central density, it is evaluated as the average density within $r=0.03~{\rm kpc}$ from the halo center, which is well resolved.

\subsection{Numerical comparisons}
\label{sec:sigmav}

\begin{figure*}

  \centering
  \includegraphics[scale=0.35]{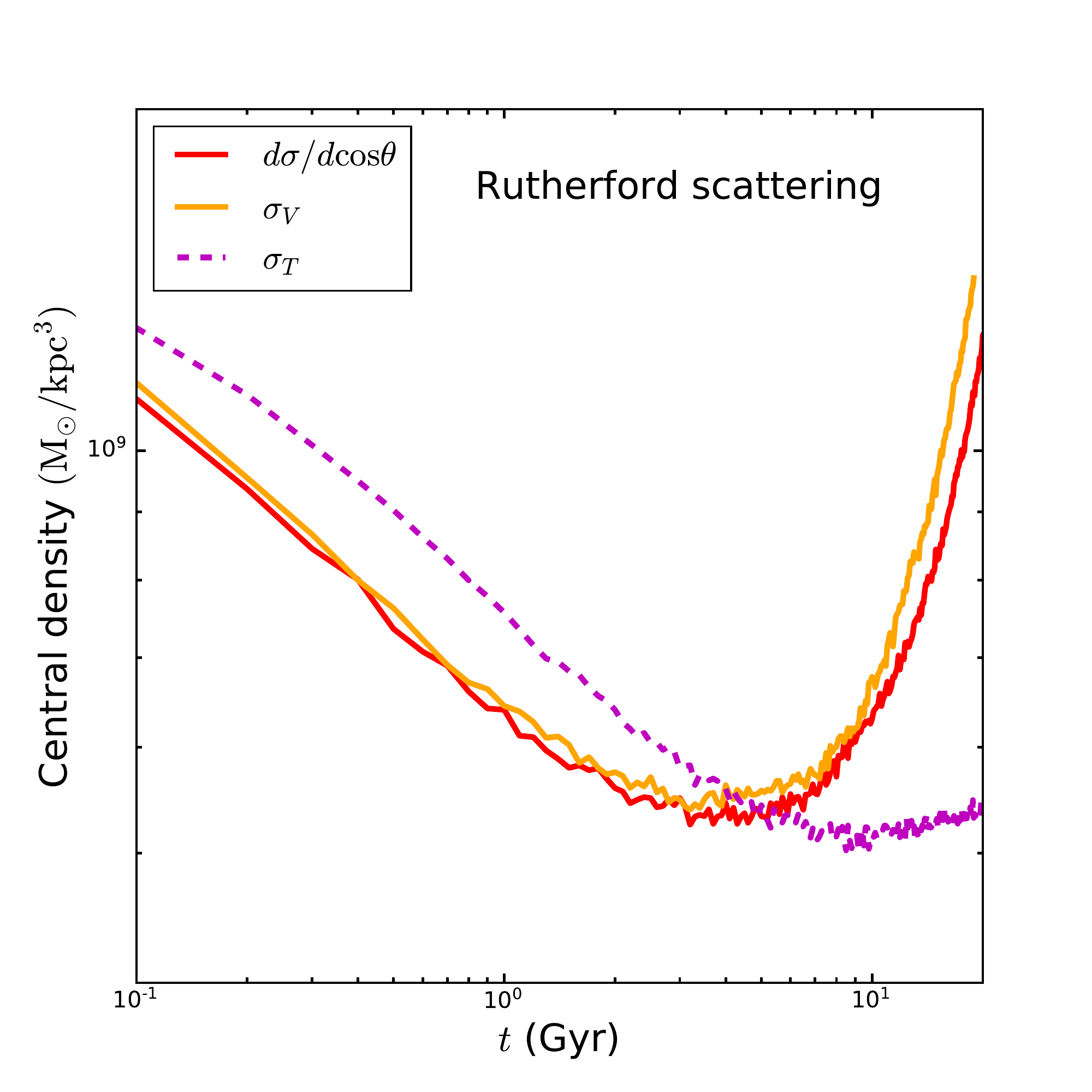}
  \includegraphics[scale=0.35]{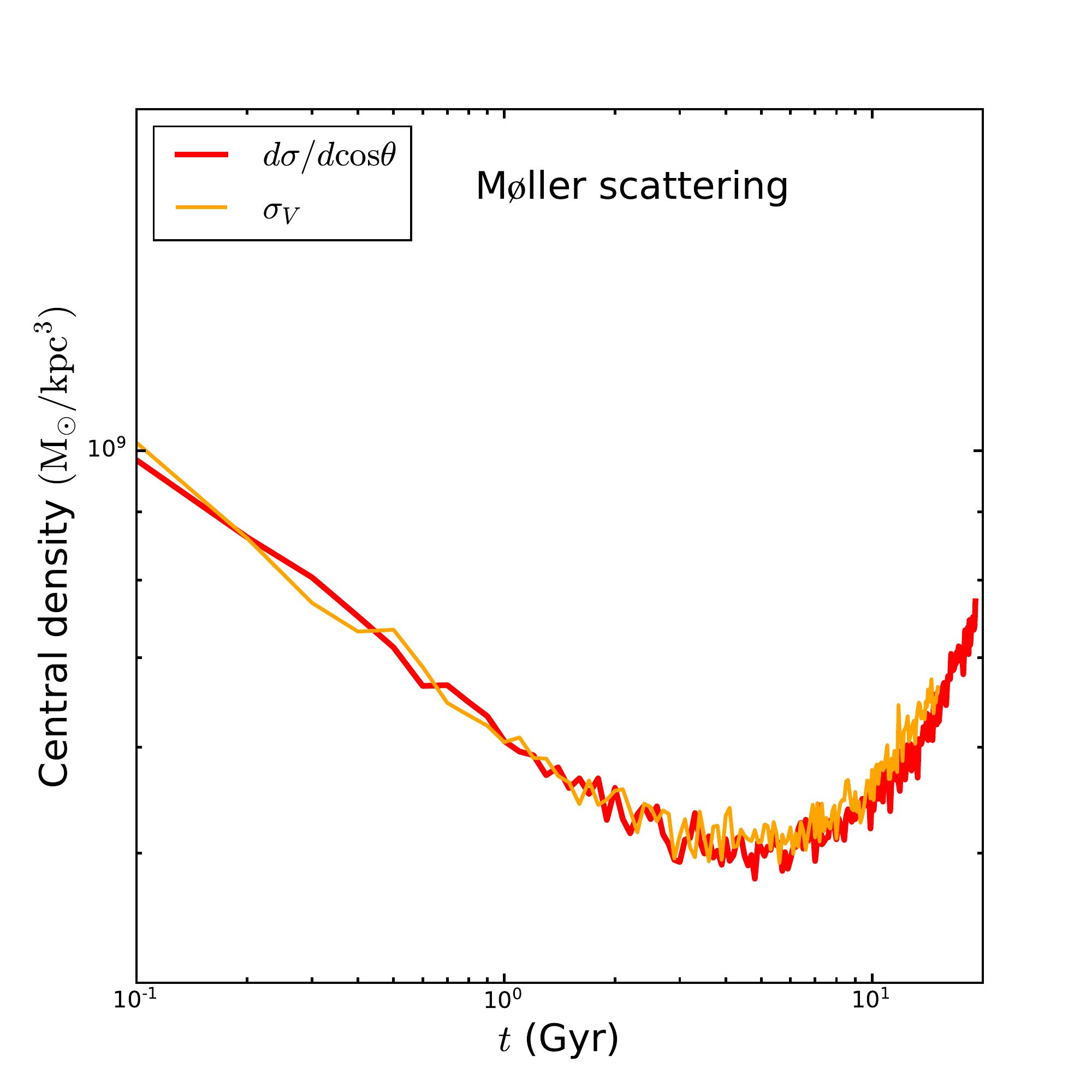}
  \caption{Gravothermal evolution of the central dark matter density for Rutherford (left panel) and \Moller (right panel) scatterings. The red and orange curves denote simulations using angular-dependent differential ($d\sigma/d\cos\theta$) and angular-independent viscosity ($\sigma_V$) cross sections, respectively. For Rutherford scattering, the magenta curve denotes simulation results using the transfer cross section $\sigma_T$.
 \label{fig:comparison}
 }     
\end{figure*}

In Fig.~\ref{fig:comparison} (left panel), we show the central dark matter density vs. evolution time for Rutherford scattering based on simulations with the differential ($d\sigma/d\cos\theta$, red), viscosity ($\sigma_V$, orange) and transfer ($\sigma_T$, magenta) cross sections. The halo first enters the core-expansion phase and the central density becomes lower. Then it evolves further into the core-collapse phase and the density increases accordingly. We see that the simulation results based on $d\sigma/d\cos\theta$ and $\sigma_V$ are very similar. Both predict almost identical central densities during the core-expansion phase and they reach minimum at $t\sim3~{\rm Gyr}$. In the core-collapse phase, the agreement in the central density is within about $10\%$ for a given snapshot. On the other hand, the transfer cross section systematically underestimates the effects of dark matter self-interactions. In particular, the central density reaches its minimum at $t\approx10~{\rm Gyr}$ in this case, a factor of $3$ longer than the actual one as found in the $d\sigma/d\cos\theta$ and $\sigma_V$ simulations.

We can understand the discrepancy as follows. From Eqs.~\ref{eq:xsrT} and~\ref{eq:xsrV}, both $\sigma_V$ and $\sigma_T$ are normalized such that $\sigma_V=\sigma_T=\sigma_{\rm tot}=\sigma_0$ for $v\ll w$, i.e., the velocity-independent limit. For $v\gg w$, $\sigma_V=6\sigma_0 [\ln(v^2/w^2)-2]/(v/w)^4$, while $\sigma_T=2\sigma_0  [\ln(v^2/w^2)-1]/(v/w)^4$. Since $w=1~{\rm km/s}$ and $v\sim10~{\rm km/s}$ for the halo, $\sigma_T/\sigma_V\sim0.4$. Thus the transfer cross section underestimates the SIDM effects in the simulations. Since the collapse timescale is inversely proportional to the size of the cross section~\cite{2002ApJ...568..475B,2011MNRAS.415.1125K,Essig:2018pzq}, the onset of the collapse is longer for $\sigma_T$. It is also useful to check the total cross section in this limit, $\sigma_{\rm tot}=\sigma_0/(v/w)^2$, which is a factor of $5$ larger than $\sigma_V$ for the case we consider. Thus in the regime where the self-interactions are strongly velocity-dependent, the total cross section does not provide a good measure, as it overestimates the actual impacts on the halo. 

Fig.~\ref{fig:comparison} (right panel) shows excellent agreement between $d\sigma/d\cos\theta$ and $\sigma_V$ simulations for \Moller scattering in both expansion and collapse phases. In this case, the viscosity cross section regulates both forward and backward scattering, it provides a good approximation for modeling \Moller scattering, where the transfer cross section cannot even be properly defined. 

Our simulations have demonstrated that the viscosity cross section in Eq.~\ref{eq:xsrD} can accurately model gravothermal evolution of the SIDM halo. For the choice of our model parameters, i.e., the relative velocity $v\sim10~{\rm km/s}$ and $w=m_\phi c/m_\chi=1~{\rm km/s}$, the scattering is extremely anisotropic, as indicated in Fig.~\ref{fig:xscomp}. Nevertheless, even in this limit, the viscosity cross section provides a good approximation for modeling differential dark matter collisions for both Rutherford and \Moller scatterings. Since $\sigma_V$ has no angular dependence, it is relatively easy to implement in N-body simulations. In addition, $\sigma_V$ itself regulates forward and backward scatterings simultaneously, and hence we can avoid dealing with those ``spurious'' events in the simulations. The success of $\sigma_V$ is due to the fact that its weighting kernel $\sin^2\theta$ characterizes the effect of heat conductivity in the halo; we will come back to this point in Sec.~\ref{sec:heat}.

\section{Thermodynamic properties of dark matter halos}

\label{sec:thermal}

We can understand the gravothermal evolution of the halo shown in Fig.~\ref{fig:comparison} from the perspective of thermodynamics. In fact, there are tremendous studies in using a semi-analytical conducting fluid model to study the evolution of SIDM halos~\cite{Balberg:2001qg,2011MNRAS.415.1125K,Pollack:2014rja,2002ApJ...568..475B,Essig:2018pzq,Nishikawa:2019lsc}. Refs.~\cite{Kaplinghat:2013xca,Kaplinghat:2015aga} apply the ideal gas law and derive the density profile of an SIDM halo in static equilibrium. We take a complementary approach by reconstructing thermodynamic quantities from our simulations directly, examining their relations and studying the implications. In this section, the reconstruction is based on the $d\sigma/d\cos\theta$ simulation for Rutherford scattering. 

\subsection{Luminosity and specific heat}

\begin{figure*}
  \centering
  \includegraphics[height=4.5cm]{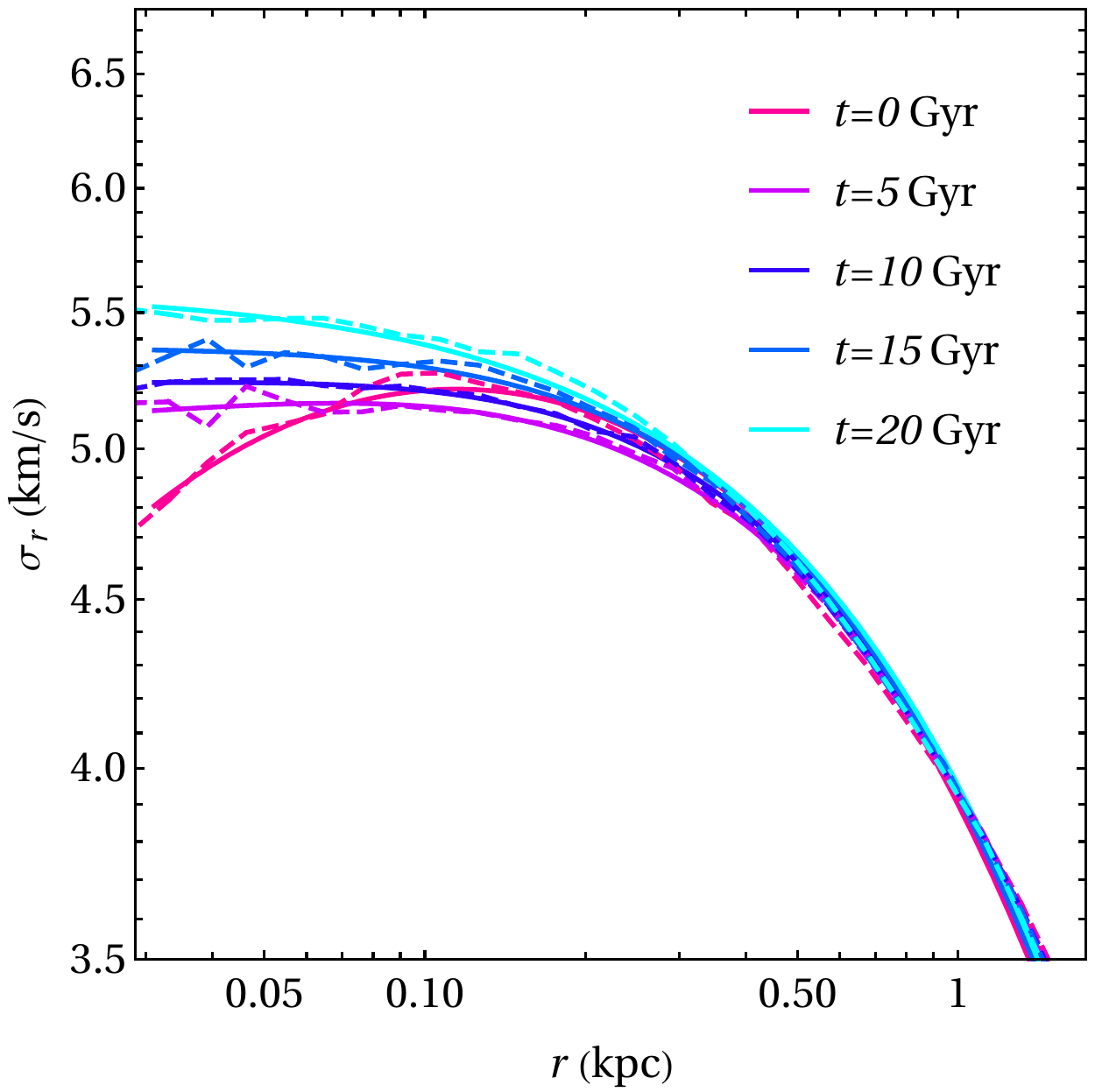}
  \includegraphics[height=4.5cm]{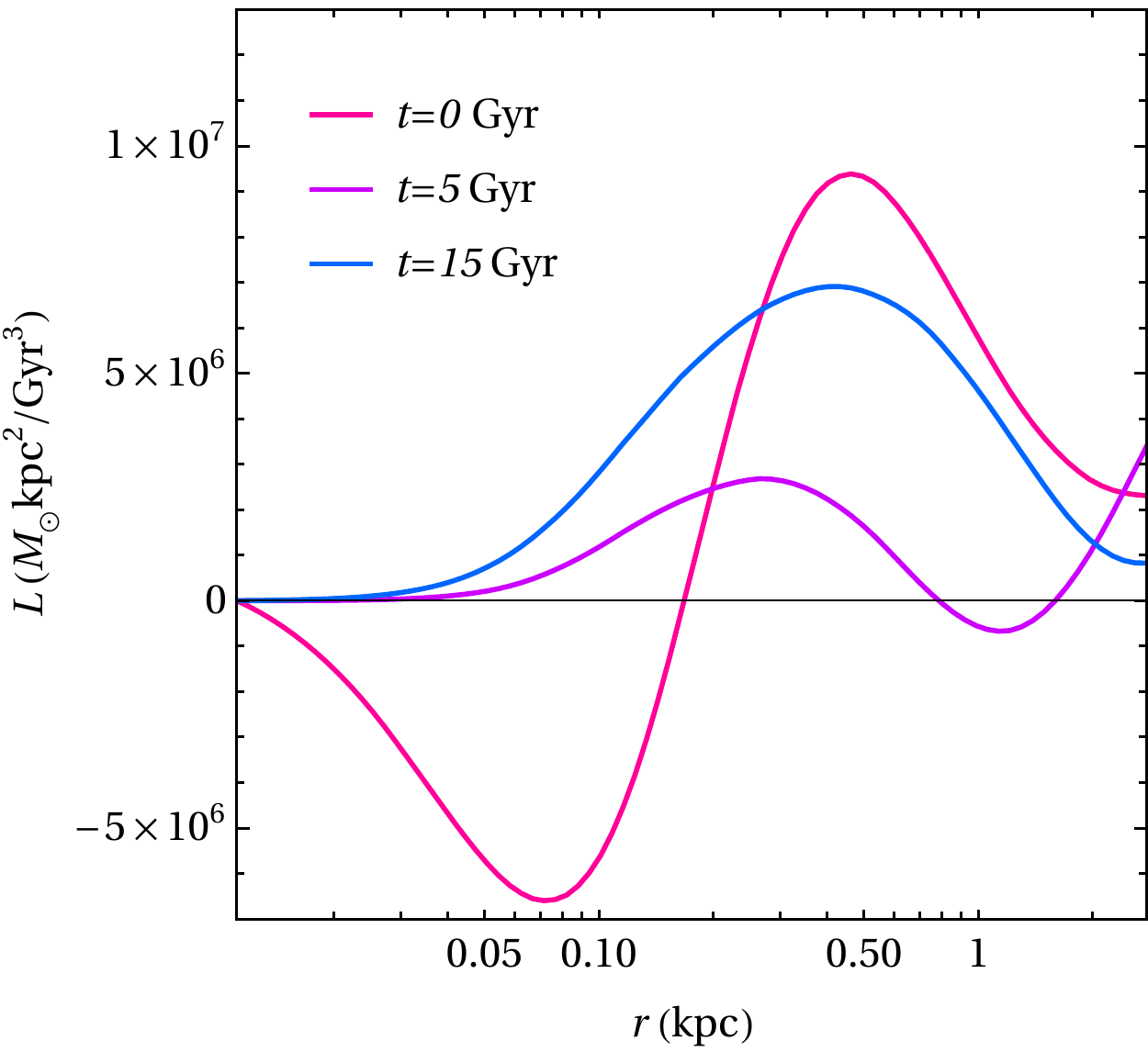}
  \includegraphics[height=4.5cm]{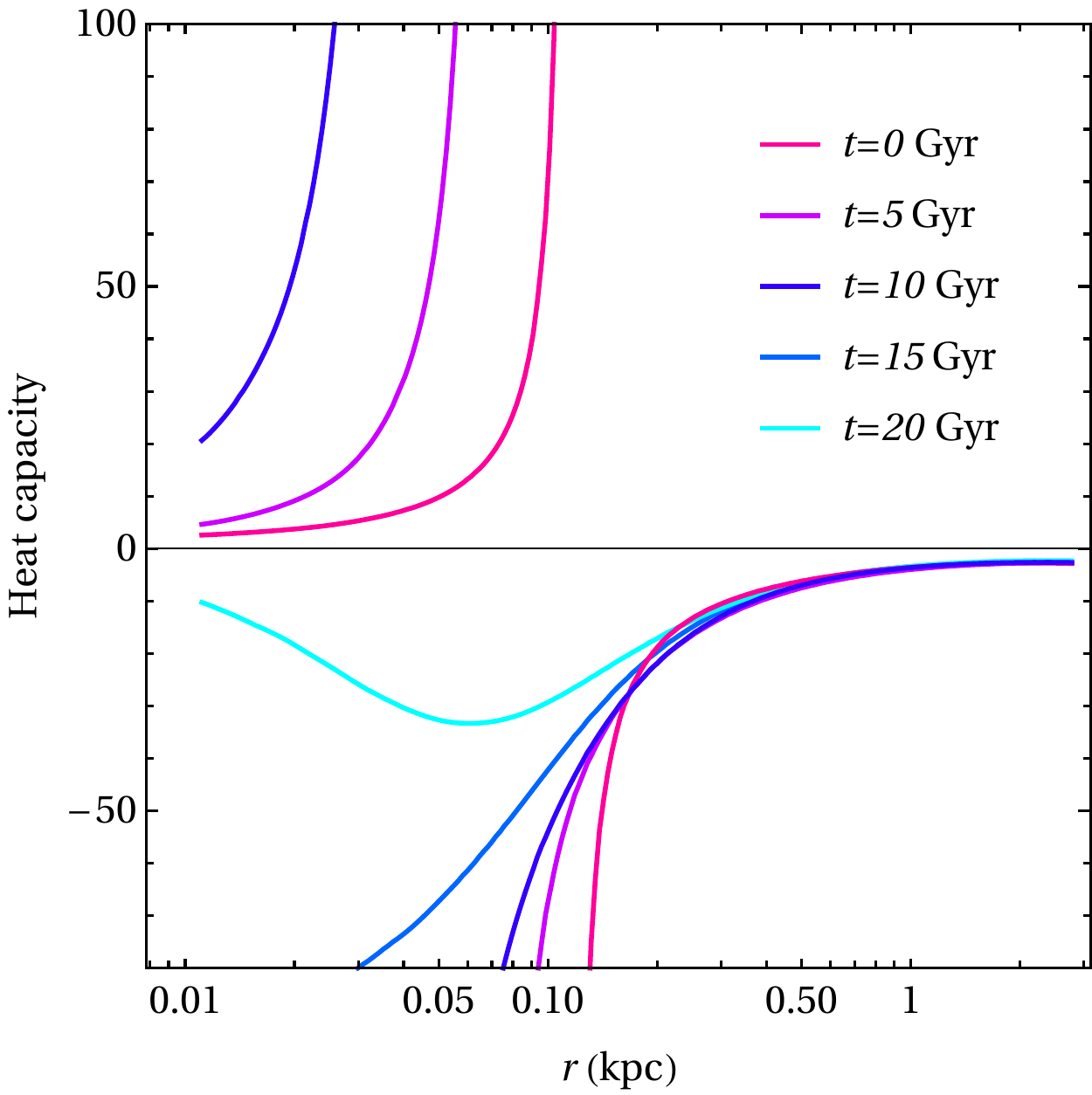}
  \caption{ {\it Left}: Profiles of the radial velocity dispersion from the $d\sigma/d\cos\theta$ simulation for Rutherford scattering (dashed) and the reconstructed one using the analytical fitting function in Eq.~\ref{eq:fitvr} (solid). {\it Middle}: Profiles of the luminosity. {\it Right}: Profiles of the specific heat capacity. 
  \label{fig:heatc}
   }
\end{figure*}

We consider luminosity and specific heat profiles of the simulated halo at different evolution times. As the first step, we fit the radial profile of the velocity dispersion of the simulated halo with the ansatz
\begin{equation}
\label{eq:fitvr}
\sigma_r(r) = \frac{a (r+b)}{(r+c)(r+d)},
\end{equation}
where the parameters $a$, $b$, $c$ and $d$ are determined by fitting to the simulated $\sigma_{r}(r)$ profile at each snapshot. Fig.~\ref{fig:heatc} (left panel) shows the fitted velocity dispersion profiles (solid) for six snapshots, as well as their corresponding simulated one (dashed). We see that the agreement is reasonably well in both expansion and collapse phases. The smooth fitting function helps us avoid numerical noises in calculating luminosity and specific heat. Our initial condition assumes that the velocity distribution is isotropic, and we set the 1D velocity dispersion to be the radial one, $\sigma_{\rm 1D}=\sigma_r$.

We calculate the specific energy as $E(r) = \frac{3}{2}\sigma_{\rm 1D}^2(r) + \Phi(r)$ for particles within a spherical shell at $r$, where the gravitational potential is given by
\begin{equation}
\Phi(r) = -4\pi G\left[\frac{1}{r}\int_0^r \rho(r') r'^2 d r' + \int_r^{\infty} \rho(r') r' d r' \right]. 
\end{equation}
In practice, we set the upper limit of $r'$ to be the virial radius of the halo $r_{200}$. For each snapshot, we interpolate the simulation results and obtain a smooth density profile. We then calculate the dimensionless specific heat capacity as $C(r)=dE/d\sigma^2_{\rm 1D}$~\cite{2008gady.book.....B}. 

In addition, we calculate the luminosity profile as 
\begin{equation}
L = -4\pi \int_0^r d r' r'^2 \rho(r') \frac{D E(r)}{D t},
\label{eq:luminosity}
\end{equation}
where $DE/Dt$ is the Lagrangian derivative of the specific energy. To compute $DE/Dt$, we search for the radius $r_{\rm M}$ such that the total enclosed mass within $r_{\rm M}$ at $t+\Delta t$ equals to the mass within $r$ at time $t$, and we take $\Delta t=1~{\rm Gyr}$. For a scalar quantity such as the specific energy $E(r,t)$, we evaluate its Lagrangian derivative as:
\begin{equation}
\frac{D E}{D t} = \frac{E(r_{\rm M},t+\Delta t)-E(r,t)}{\Delta t}. 
\label{eq:time}
\end{equation}

In Fig.~\ref{fig:heatc}, we show radial profiles of the luminosity (middle panel) and the specific heat capacity (right panel) at different evolution times. At the early stage $t\sim0\textup{--}5~{\rm Gyr}$, the luminosity is negative in the inner region $r\lesssim r_s\approx0.11~{\rm kpc}$, indicating that that energy is transferred inwards. Since the heat capacity is {\it positive} in the region during the time window, the inner halo is heated up. For $t\sim5~{\rm Gyr}$, the inner luminosity is vanishing and the radial gradient of $\sigma_{\rm 1D}$ becomes small, heat conduction is suppressed, while the heat capacity is still positive. Consider $t\gtrsim15~{\rm Gyr}$, both gradients of the velocity dispersion and heat capacity are negative for the whole halo, but the luminosity is positive. At this stage, the halo is deeply in the collapse phase. The central halo becomes hot and its density increases continuously as dark matter self-interactions pump the heat outwards. 

It is interesting to note that the timescale for the gradient of the inner velocity dispersion becoming negative ($t\gtrsim10~{\rm Gyr}$) is much longer than that for forming a density core. As shown Fig.~\ref{fig:comparison} (left panel, red), the density core forms quickly around $t\sim1~{\rm Gyr}$ and it remains rather stable before the collapse starts. Since $\Phi(r)$ increases monotonically with $r$ and a negative specific heat capacity requires $d\sigma(r)/d r<0$, which takes a long evolution time to achieve. In addition, although the heat capacity of the inner halo evolves and changes from positive to negative values, the heat capacity of the outer halo rarely evolves and remains negative. This is because for the outer halo, the self-scattering rate is low and the velocity dispersion has a negative radial gradient.

\subsection{Fluid description}

For SIDM, the phase space density of dark matter particles $f$ can be modeled by the Boltzmann equation with a collision term $C[f]$,
\begin{equation}
\label{eq:boltzmann0}
\frac{d f}{d t} = \frac{\partial f}{\partial t} + \mathbf{v}\cdot \mathbf{\nabla} f - \mathbf{\nabla} \Phi \frac{\partial f}{\partial \mathbf{v}} = C[f], 
\end{equation}
together with Poisson's equation $\mathbf{\nabla}^2 \Phi = 4\pi G \rho = 4\pi G \int d^3 v f$. 

When dark matter particles collide frequently, one may close the first three moment equations of the Boltzmann equation in Eq.~\ref{eq:boltzmann0} by introducing thermal conductivity and viscosity~\cite{lifshitz1995physical}. This leads to a fluid approximation, which has been used extensively for modeling the evolution of SIDM halos. However, it is not obvious that the fluid description applies to self-gravitating systems, like dark matter halos, as many results from statistical mechanics are no longer valid, see the discussion in box 7.1 of Ref.~\cite{2008gady.book.....B}. For example, energy is not an extensive quantity for a self-gravitating system because the contribution from distant particles is important, the microcanonical probability distribution cannot be defined properly as its energy hypersurface is unbounded, and the heat capacity is negative as the total energy is negative~\cite{2008gady.book.....B}.  We use our simulation results to explicitly test the moment equations and confirm the validity of the fluid description. 

The zeroth moment gives rise to the continuity equation $dM(r)/dr=4\pi r^2\rho$, which is trivially satisfied in N-body simulations. From the first moment of the Boltzmann equation, we get 
\begin{equation}
\label{eq:pbalance}
\mathbf{\nabla}(\rho \sigma^2_{\rm 1D}) = - \rho \mathbf{\nabla} \Phi.
\end{equation}
For a system in hydrostatic equilibrium, the ``buoyancy'' force generated by the gradient of the pressure $\rho\sigma^2_{\rm 1D}$ is balanced by gravity. The second moment describes the heat transport, relating the gradient of the luminosity to the change rate of the entropy as 
\begin{equation}
\frac{1}{4\pi}\frac{\partial L}{\partial r}=-\rho\sigma^2_{\rm 1D}\frac{Ds}{Dt},
\label{eq:luminosity2}
\end{equation}
where $s=\ln\sigma^3_{\rm 1D}/\rho$ is the specific entropy~\cite{2008gady.book.....B}, and $D/Dt$ is the Lagrangian derivative. In addition, $L/4\pi r^2=-\kappa\nabla T$, where $\kappa$ is the heat conductivity and $T$ is the temperature. For a system that reaches {\it local} equilibrium, $T=m_\chi\sigma^2_{\rm 1D}$. Combining Eqs.~\ref{eq:luminosity} and~\ref{eq:luminosity2}, we have 
\begin{equation}
\frac{1}{\sigma^2_{\rm 1D}} \frac{D E}{D t} = \frac{D s}{D t}=\frac{D}{Dt}\ln\frac{\sigma^3_{\rm 1D}}{\rho}. 
\label{eq:transx}
\end{equation}
We use our simulation results to explicitly examine the conditions given in Eqs.~\ref{eq:pbalance} and~\ref{eq:transx}. 

\begin{figure*}
  \centering
  \includegraphics[height=6.8cm]{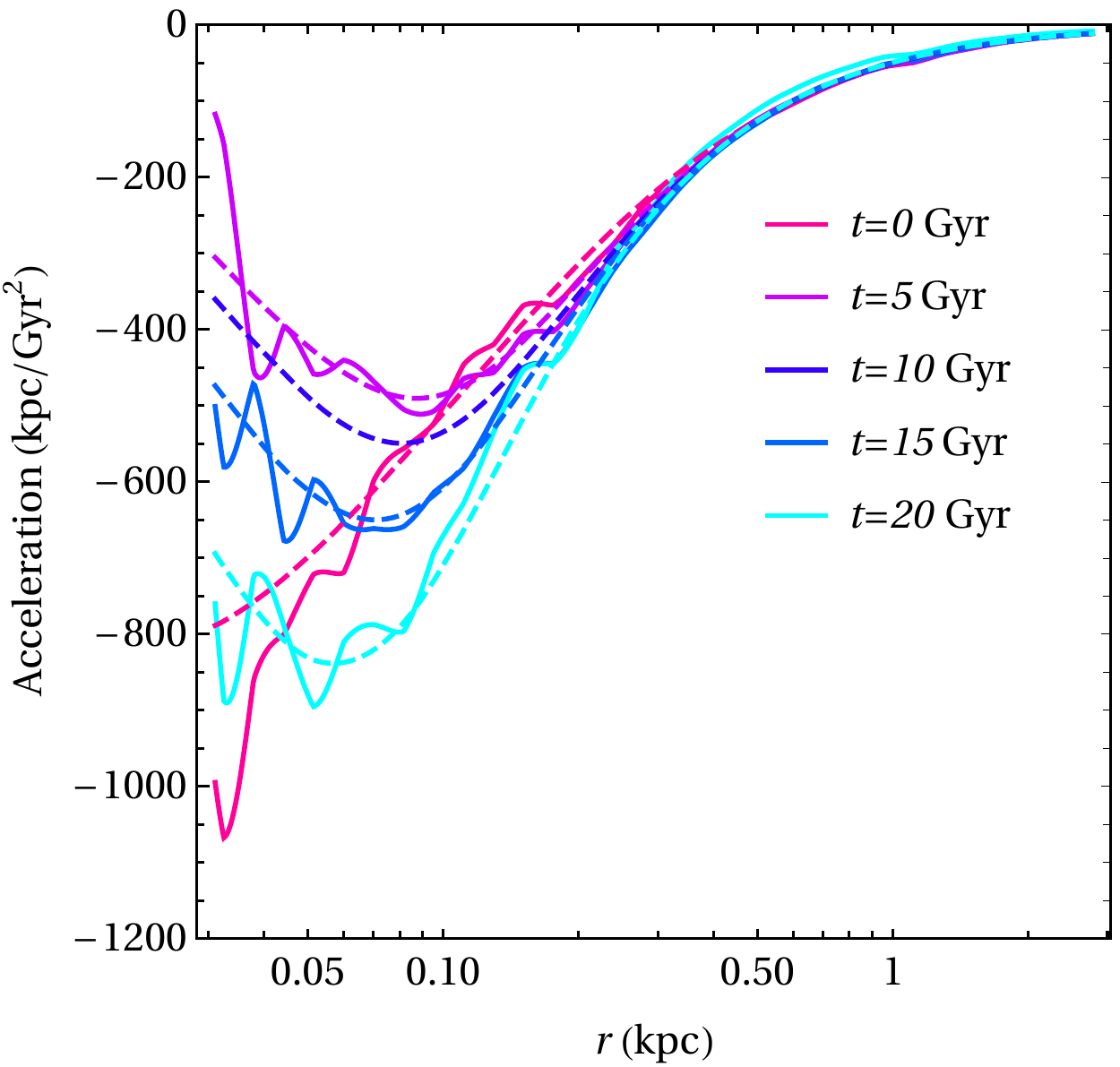}
  \includegraphics[height=6.7cm]{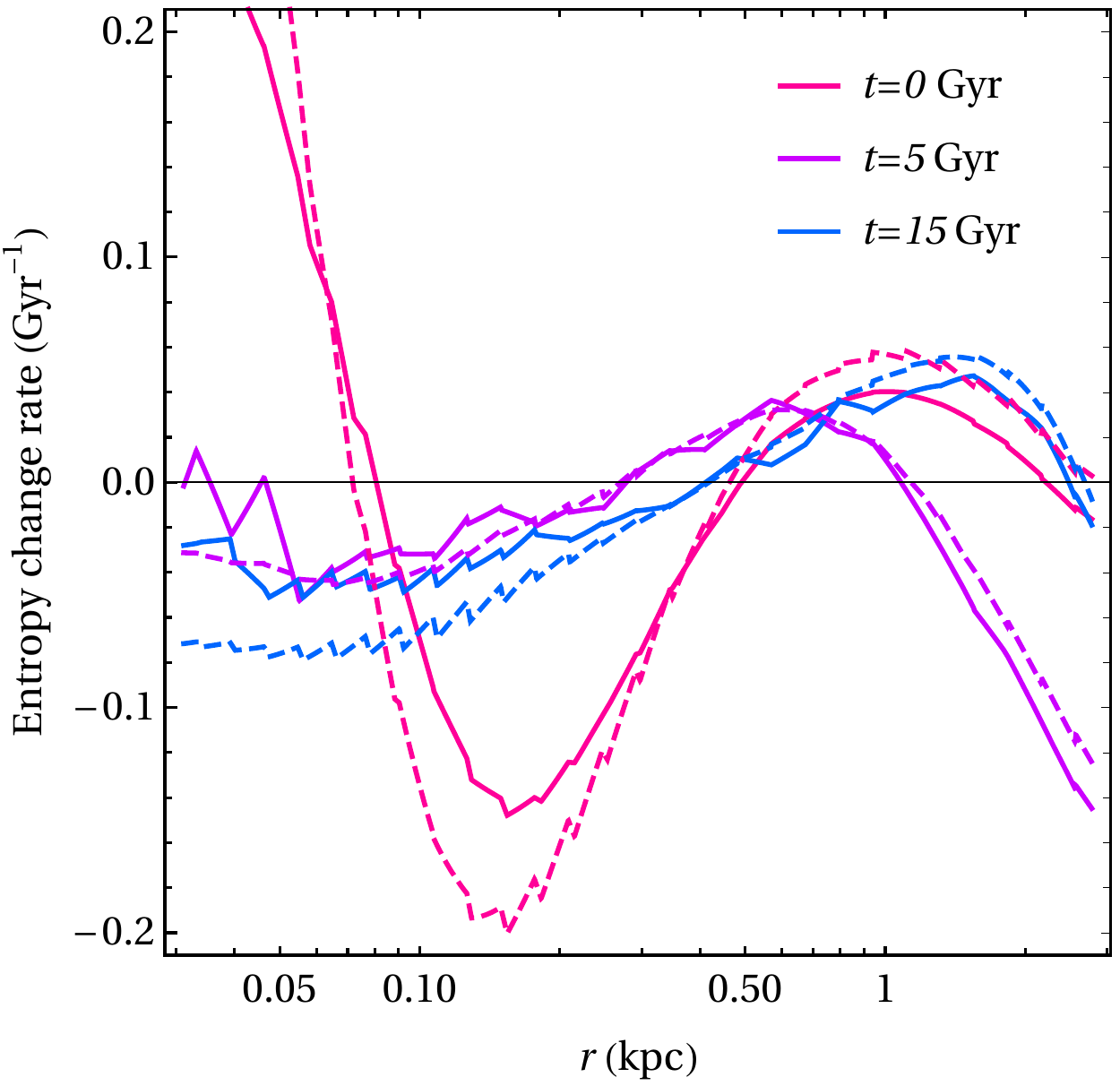}
  \caption{\label{fig:checktrans} 
{\it Left:} The profiles of centripetal acceleration due to gravity (dashed) and buoyancy (solid) from the simulated halo. {\it Right:} The profiles of $Ds/Dt$ (solid) and $(1/\sigma^2_{\rm 1D})DE/Dt$ (dashed), see Eq.~\ref{eq:transx}. 
 }
\end{figure*}

Fig.~\ref{fig:checktrans} (left panel) shows the centripetal acceleration due to gravity $d\Phi(r)/dr$ (dashed) and buoyancy acceleration $-(1/\rho)d(\rho\sigma^2_{\rm 1D})/dr$ (solid) for five snapshots. We see that the quasi-equilibrium condition Eq.~\ref{eq:pbalance} is well satisfied for different stages of gravothermal evolution of the SIDM halo, even in the collapse phase. Fig.~\ref{fig:checktrans} (right panel) shows the radial profiles of $Ds/Dt$ (solid) and $(1/\sigma^2_{\rm 1D})DE/Dt$ (dashed) for three representative snapshots, and they match reasonably well, indicating the condition~\ref{eq:transx} is satisfied for the simulated halo. The small oscillatory features in both panels are numerical noises in computing the derivative of the density profile from the simulations.

\subsection{Heat conductivity}
\label{sec:heat}

As discussed, the luminosity is related to the gradient of temperature as $L/4\pi r^2=-\kappa\nabla T$. In kinetic theory, the heat conductivity $\kappa$ is calculated in the following way, see, e.g.,~\cite{lifshitz1995physical}. The collisions lead to a small deviation from a local equilibrium Maxwellian distribution, and we can parametrize the distribution perturbation as $\delta f = (\bar{f}/T) \mathbf{g}\cdot\mathbf{\nabla} T$ and compute the conductivity as~\cite{lifshitz1995physical}
\begin{eqnarray}
\kappa &=& -\frac{1}{3 T} \int d^3v \frac{v^2}{2} \mathbf{v}\cdot \mathbf{g} \bar{f}, 
\end{eqnarray}
where the vector function $\mathbf{g}$ can be evaluated as a series expansion of the associated Laguerre polynomials. 
Based on the first non-vanishing contribution to $\mathbf{g}$, one obtains~\cite{lifshitz1995physical}
\begin{equation}
\label{eq:kappashort}
\kappa = \frac{75}{16}\left[ \frac{1}{4 \sqrt{2\pi}(\sqrt{2}\sigma_{\rm 1D})^9} \int d v d \cos\theta \exp\left[{-\frac{v^2}{2 (\sqrt{2}\sigma_{\rm 1D})^2}}\right] v^7 \sin^2\theta \frac{d\sigma}{d\cos\theta}\right]^{-1}, 
\end{equation}
where $v$ is the relative velocity between two initial states and $\theta$ is the scattering angle.

The conductivity calculated in Eq.~\ref{eq:kappashort} is valid in the short mean-free-path regime, and hence we will denote it as $\kappa_{\rm smfp}$. For a constant cross section $\sigma$, $\kappa_{\rm smfp}\approx 2.1\sigma_{\rm 1D}/\sigma$. 
In this regime, the length scale of heat conduction is characterized by the mean free path between two consequential interactions $\lambda=1/(n\sigma)$, where $n$ is the number density of particles and $\sigma$ is the self-scattering cross section. 
However, it is well known that for an SIDM halo, it is in the long-mean-free-path regime for the majority of its evolution history. Only at late stages of gravothermal collapse, the central halo may have frequent enough dark matter scattering and $\kappa_{\rm smfp}$ applies. In the long-mean-free-path regime, the conducting fluid model introduces an empirical conductivity $k_{\rm lmfp}\approx0.27\beta n\sigma^3_{\rm 1D}\sigma/(Gm_\chi)$ for a constant cross section~\cite{Lynden-Bell:1968eqn,Balberg:2001qg}, where the numerical factor $\beta$ can be determined by calibrating to N-body simulations. Studies show that $\beta\approx0.75$ and $0.60$ for isolated and cosmological simulations with a constant cross section~\cite{Essig:2018pzq}.

We use our simulation results to {\it directly} estimate the conductivity $\kappa$ in the long mean-free-path regime and compare it with the empirical one. Taking the relations $L/4\pi r^2=-\kappa\nabla T$ and $T=m\sigma^2_{\rm 1D}$, we have $\kappa m_\chi=L/[4\pi r^2d\sigma^2_{\rm 1D}/d r]$. From the luminosity profile reconstructed from the simulated halo as shown in Fig.~\ref{fig:heatc} (middle panel), we estimate the conductivity at $r=0.5~{\rm kpc}$ as
\begin{eqnarray}
\nonumber (\kappa_{\rm }m_\chi)_{\rm est} &\approx& \frac{L}{4\pi r^2 2 \sigma_{\rm 1D} (d\sigma_{\rm1D}/d r) } \\ 
\nonumber & \approx& \frac{5\times 10^6 ~\rm M_{\odot} kpc^2 Gyr^{-3}}{4\pi (0.5~{\rm kpc})^2 (2\times 5~{\rm kpc/Gyr}) (3~{\rm kpc/Gyr}/(1~{\rm kpc})) } 
  \approx 0.5\times 10^5 ~\rm{M_{\odot}/kpc/Gyr}, 
\end{eqnarray}
which agrees the empirical conductivity evaluated at $r=0.5~ \rm kpc$
\begin{equation}
\label{eq:kappalong}
\kappa_{\rm lmfp}m_\chi = 0.27\times 0.75 \rho \sigma^3_{\rm 1D} \frac{\sigma_V}{G} \approx 10^5 ~\rm M_{\odot}/kpc/Gyr,
\end{equation}
where we take the viscosity cross section $\sigma_V$ in evaluating $\kappa_{\rm lmfp}m_\chi$. Note that the reconstructed conductivity depends both radius and time, and we have taken characteristic values for the variables $L$, $\sigma_{\rm 1D}$ and $d\sigma_{\rm 1D}/dr$ in estimating $(\kappa_{\rm }m_\chi)_{\rm est}$. Thus the small difference between $(\kappa_{\rm }m_{\chi})_{\rm est}$ and $\kappa_{\rm lmfp}m_{\chi}$ is not surprising. For comparison, we further calculate $\kappa_{\rm smfp}m_\chi$ at $r=0.5~{\rm kpc}$ directly using Eq.~\ref{eq:kappashort} and find 
\begin{equation}
\kappa_{\rm smfp}m_\chi\approx 5\times 10^9~ \rm M_{\odot} /kpc /Gyr,
\end{equation}
being $\sim4\textup{--}5$ orders of magnitudes larger. Thus our simulations directly confirm the validity of the empirical conductivity for the long-mean-free-path regime. 

\section{The constant effective cross section}

\label{sec:sigmaeff}

In Sec.~\ref{sec:sigmav}, we have shown that the viscosity cross section $\sigma_V$ provides a good approximation to model differential dark matter self-scattering. The viscosity cross section does not have an explicit angular dependence and it regulates spurious forward and backward scatterings. However, $\sigma_V$ is still velocity-dependent. In this section, we propose a constant effective cross section, which integrates over a characteristic velocity dispersion for a given halo. Our simulations show the constant effective cross section provides an approximation to differential self-scattering for most of the halo evolution.

\subsection{The effective cross section and its validation}

\label{sec:eff}

As we discussed, for an SIDM halo, the scattering is mostly in the long-mean-free-path regime, and hence the heat conductivity introduced in Eq.~\ref{eq:kappashort}, which is valid in the short-mean-free-path regime, cannot be used in the conducting fluid model for studying the evolution of the entire halo. However, heat conduction is based on dark matter collisions, which occur locally in both short- and long-mean-free-path regimes. The fluid model uses $\kappa_{\rm lmfp}$ to incorporate effects of orbital evolution of a particle after the collision~\cite{Lynden-Bell:1968eqn}, while N-body simulations automatically take them into account by construction. We expect the heat conductivity in Eq.~\ref{eq:kappashort} provides a good approximation for capturing local heat transport properties of dark matter self-interactions. 

We first introduce a local conductivity cross section motivated by Eq.~\ref{eq:kappashort},
\begin{equation}
\label{eq:xsk}
\sigma_{\kappa}(r) = \frac{2 \int v^2d v d \cos\theta \frac{d \sigma}{d \cos\theta} \sin^2\theta v^5 \exp \left[-\frac{v^2}{4\sigma_{\rm 1D}^2(r)}\right]}{\int v^2d v d \cos\theta \sin^2\theta v^5  \exp \left[-\frac{v^2}{4\sigma_{\rm 1D}^2(r)}\right]},
\end{equation}
where the differential cross section $d\sigma/d\cos\theta$ is both angular and velocity-dependent and the $\sigma_{\rm 1D}(r)$ is the radial 1D velocity dispersion profile. The normalization is chosen such that a cross section with no angular- and velocity-dependence integrates to give the total cross section. In practice, it will be more convenient if there is a single characteristic velocity dispersion for a given halo, such that we can remove the radial dependence in $\sigma_{\kappa}$. The conductivity cross section introduced in Eq.~\ref{eq:xsk}, the weighting kernel for the angular dependence is $\sin^2\theta$, which is the same as the viscosity cross section defined in Eq.~\ref{eq:xsrD}. This is not surprising because viscosity of the fluid is also related to $\sigma_\kappa$~\cite{lifshitz1995physical}, which motivates the definition in Eq.~\ref{eq:xsrD}.

Suppose such a characteristic velocity dispersion $\sigma^{\rm eff}_{\rm 1D}$ exists. After replacing $\sigma_{\rm 1D}(r)$ with $\sigma^{\rm eff}_{\rm 1D}$, we perform the integration for the denominator and obtain an effective cross section:
\begin{eqnarray}
\label{eq:eff}
\sigma_{\rm eff} = \frac{1}{512 (\sigma^{\rm eff}_{\rm 1D})^8 } \int v^2d v d\cos\theta \frac{d \sigma}{d\cos\theta} v^5 \sin^2\theta \exp \left[-\frac{v^2}{4(\sigma^{\rm eff}_{\rm 1D})^2}\right].
\end{eqnarray}

We first consider a trial case with $\sigma/m_\chi=10~{\rm cm^2/g}$, corresponding to $\sigma^{\rm eff}_{\rm 1D} = V_{\rm max}/\sqrt{3}$. For the BM2 halo, $V_{\rm max}/\sqrt{3}\approx4.6~{\rm km/s}$, which is roughly an averaged value of the velocity dispersion within $0.1~{\rm kpc}$ of the NFW initial halo. A more optimal choice is to take the 1D central velocity dispersion when the halo reaches the maximal core expansion at which the central density is lowest. In this case, $\sigma^{\rm eff}_{\rm 1D}\approx5.14~{\rm km/s}$ at $t\approx4~{\rm Gyr}$ from the $d\sigma/d\cos\theta$ simulation (BM2), and we have $\sigma^{\rm eff}_{\rm 1D} \approx1.1V_{\rm max}/\sqrt{3}\approx0.64 V_{\rm max}$, resulting in an effective cross section of $\sigma_{\rm eff}/m_\chi=7.1~{\rm cm^2/g}$. Ref.~\cite{Outmezguine:2022bhq} used a semi-analytical fluid model and found that the dispersion is $0.64 V_{\rm max}$ when an SIDM halo has the lowest central density for different choices of model parameters. Thus $\sigma^{\rm eff}_{\rm 1D}\approx0.64 V_{\rm max}$ could hold universally.

\begin{figure}[t]
  \centering
  \includegraphics[width=8.2cm]{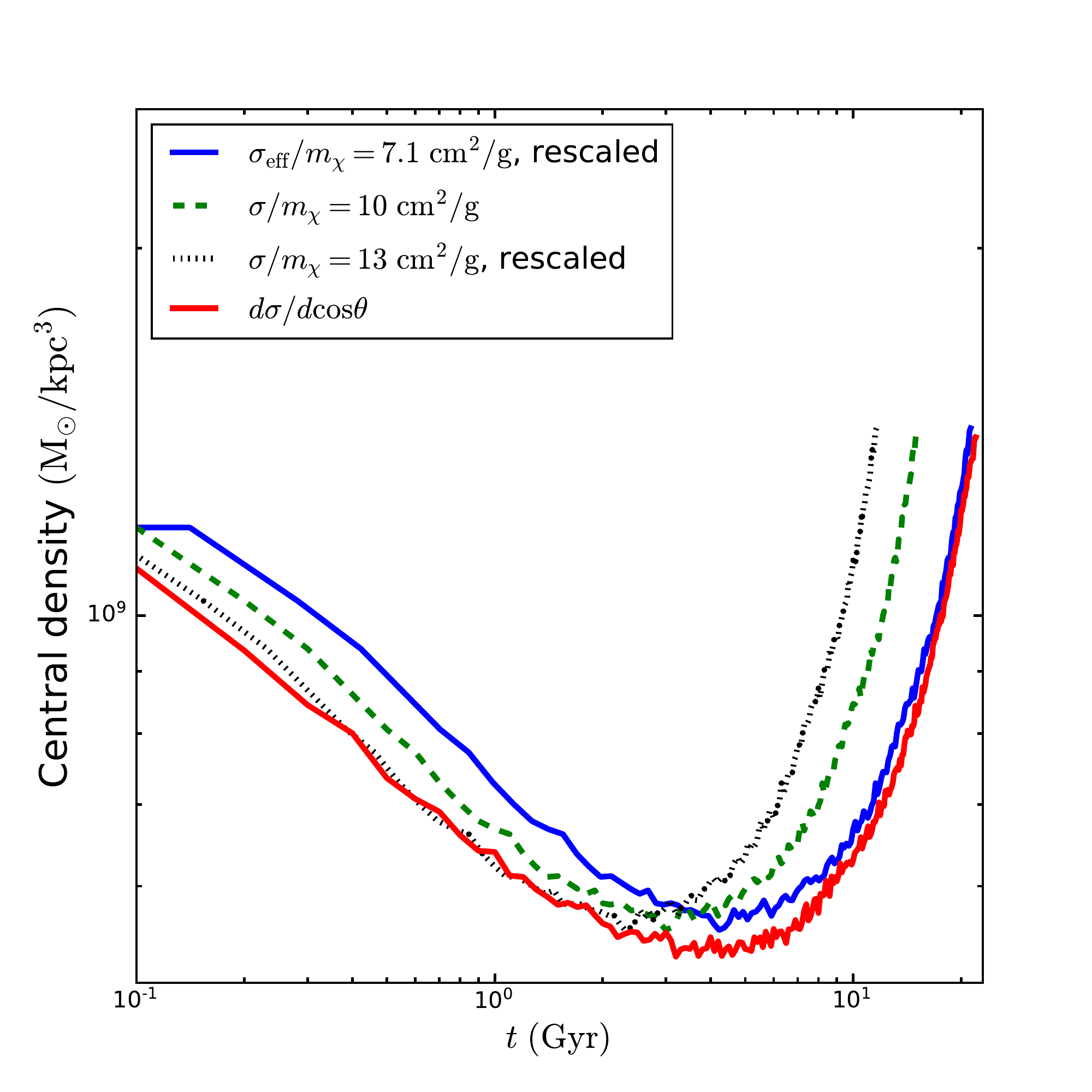}
    \caption{\label{fig:sigmaeffective} Gravothermal evolution of central dark matter densities for Rutherford scattering from simulations with a constant effective cross section $\sigma_{\rm eff}/m_\chi=7.1~{\rm cm^2/g}$ (solid blue) and differential cross section $d\sigma/d\cos\theta$ (solid red) as in Fig.~\ref{fig:comparison} (left panel). For comparison, the results with $\sigma/m_{\chi}=10~{\rm cm^2/g}$ (green dashed) and $13~{\rm cm^2/g}$ (dotted black) are also shown.}
  
\end{figure}

Fig.~\ref{fig:sigmaeffective} shows the central density vs. evolution time for $\sigma/m_\chi=10~{\rm cm^2/g}$ (dashed green), $\sigma/m_\chi=7.1~{\rm cm^2/g}$ (solid blue), compared to the $d\sigma/d\cos\theta$ simulation (solid red). We see that overall $\sigma/m_\chi=10~{\rm cm^2/g}$ is slightly too large, but $\sigma/m_\chi=7.1~{\rm cm^2/g}$ well captures the halo evolution in the core-collapse regime. At earlier stages, the effective cross section underestimates the self-scattering effect, and one needs to consider a larger cross section $\sigma/m_{\chi}=13~\rm cm^2/g$ for a precise match (dotted black). For the results with the constant cross sections shown in Fig.~\ref{fig:sigmaeffective}, only the $\sigma/m_\chi=10~{\rm cm^2/g}$ case is based on N-body simulations, while the two others are obtained by using a rescaling method based on the relation $t\propto (\sigma/m_{\chi})^{-1}$, see Appendix~\ref{app:conv} for the justification of the method.

\begin{figure}[t]
  \centering
  \includegraphics[scale=0.55]{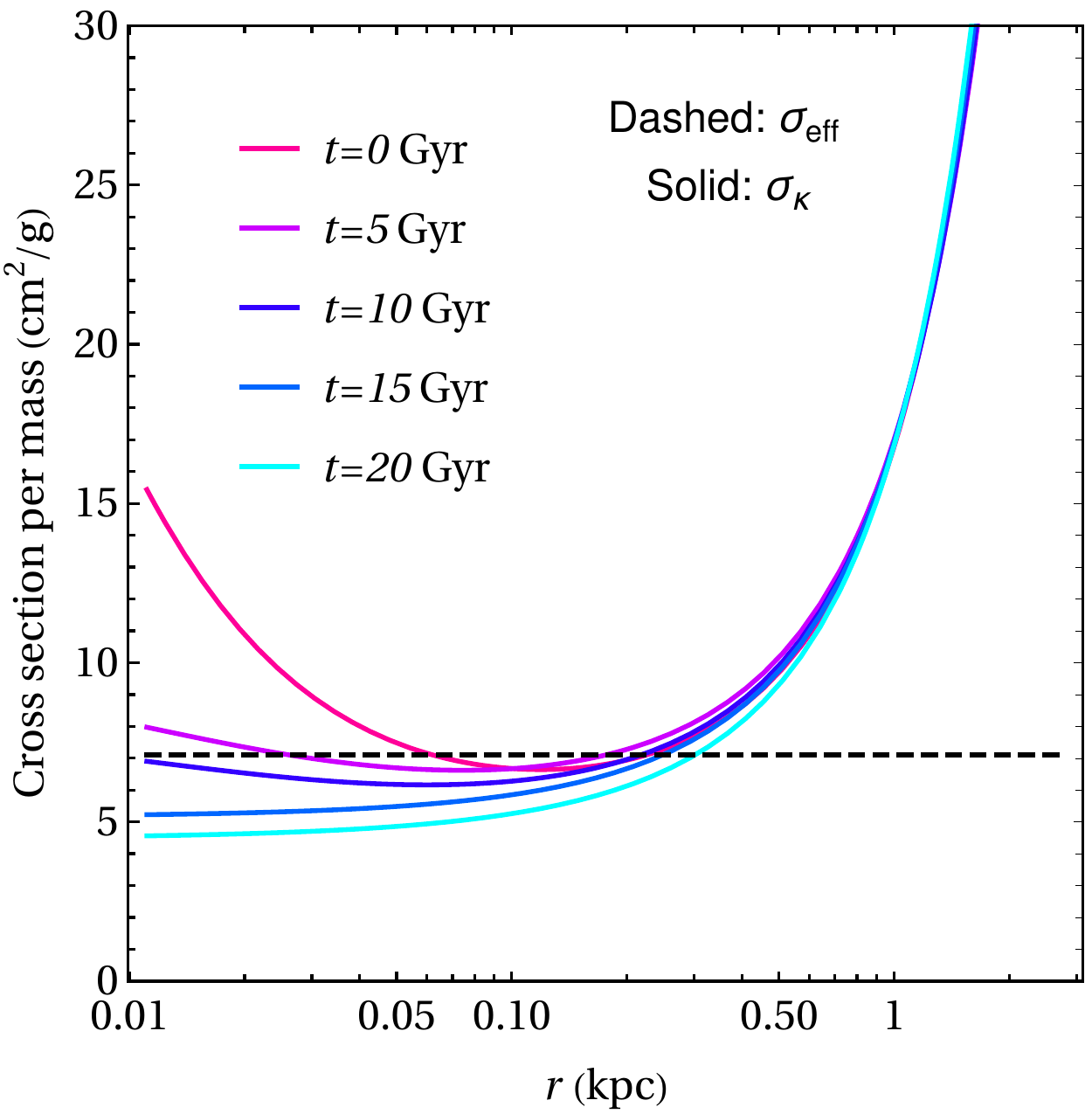}~~~
   \includegraphics[scale=0.58]{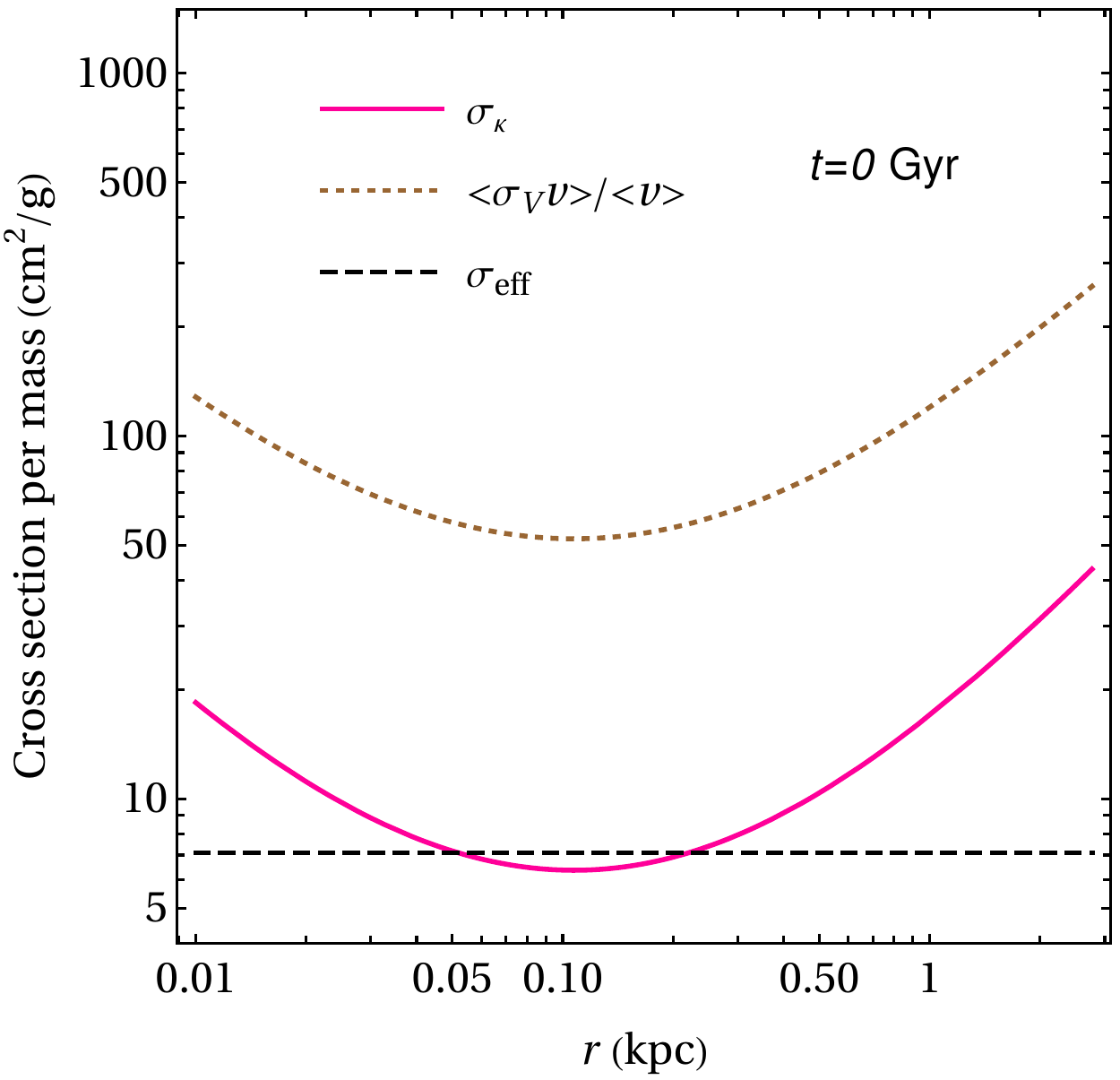}
  \caption{{\it Left:} The profiles of the local conductivity cross section $\sigma_\kappa/m_\chi$ (solid) in Eq.~\ref{eq:xsk} at different evolution times for the simulated halo and the effective cross section $\sigma_{\rm eff}/m_\chi=7.1~{\rm cm^2/g}$ calculated using Eq.~\ref{eq:eff}  (dashed). {\it Right:} The profiles of  $\left<\sigma_V v\right>/(\left<v\right>m_\chi)$ (dotted) and $\sigma_\kappa/m_\chi$ (solid) evaluated for the initial halo $t=0~{\rm Gyr}$, as well as $\sigma_{\rm eff}/m_\chi=7.1~{\rm cm^2/g}$ (dashed).
\label{fig:sigmak} 
}
\end{figure}

Fig.~\ref{fig:sigmak} (left panel) shows the conductivity cross section in Eq.~\ref{eq:xsk} at different evolution times (solid), compared to the constant effective cross section calculated using Eq.~\ref{eq:eff}, i.e., $\sigma_{\rm eff}/m_\chi=7.1~{\rm cm^2/g}$ (dashed). At early stages $t\ll5~{\rm Gyr}$, $\sigma_{\kappa}/m_\chi>\sigma_{\rm eff}/m_\chi$ at the center because $\sigma_\kappa/m_\chi$ is enhanced as the velocity dispersion decreases towards the center of an NFW halo. Thus the core formation is faster for the velocity-dependent differential cross section, as shown in Fig.~\ref{fig:sigmaeffective}. 
The constant cross section $\sigma/m_{\chi}=13~\rm cm^2/g$ is a better approximation to $\sigma_\kappa/m_\chi$ for $t<2~$Gyr, as indicated in Fig.~\ref{fig:sigmaeffective}. At later stages, especially in the collapse phase $t>5~{\rm Gyr}$, the velocity dispersion increases in the central region and $\sigma_{\kappa}$ becomes suppressed and smaller than $\sigma_{\rm eff}$ gradually. Over the evolution history of the halo up to $20~{\rm Gyr}$, $\sigma_{\rm eff}/m_\chi=7.1~\rm cm^2/g$ provides a reasonable approximation to $\sigma_{\kappa}/m_\chi$ in the inner regions $r\lesssim0.5~{\rm kpc}$.

The success of the effective cross section relies on its angular and velocity weighting kernels, which are $\sin^2\theta$ and $v^5$, respectively, see Eqs.~\ref{eq:eff} and~\ref{eq:xsk}. Note the $v^2$ factor belongs to the integration measure. Alternatively, one may consider weighting the cross section with kennels of $\sin^2\theta$ and $v$, resulting a normalized cross section $\left<\sigma_V v\right>/\left<v\right>$, where $\left<...\right>$ represents thermal averaging. The factor $v$ comes from a conventional estimate of the collision rate. Fig.~\ref{fig:sigmak} (right panel) show profiles of $\left<\sigma_V v\right>/(\left<v\right>m_\chi)$ and $\sigma_{\kappa}/m_\chi$, assuming $\sigma_{\rm 1D}(r)$ for the BM2 initial halo. We see $\left<\sigma_V v\right>/(\left<v\right>m_\chi)$ is a factor of $\sim5$ larger than $\sigma_{\kappa}/m_\chi$ and $\sigma_{\rm eff}/m_\chi$, too large to be consistent with the $d\sigma/d\cos\theta$ simulation.

\subsection{Gravothermal collapse: an extreme test}
\label{sub:diff}

\begin{figure}[t]
  \centering
  \includegraphics[width=7.2cm]{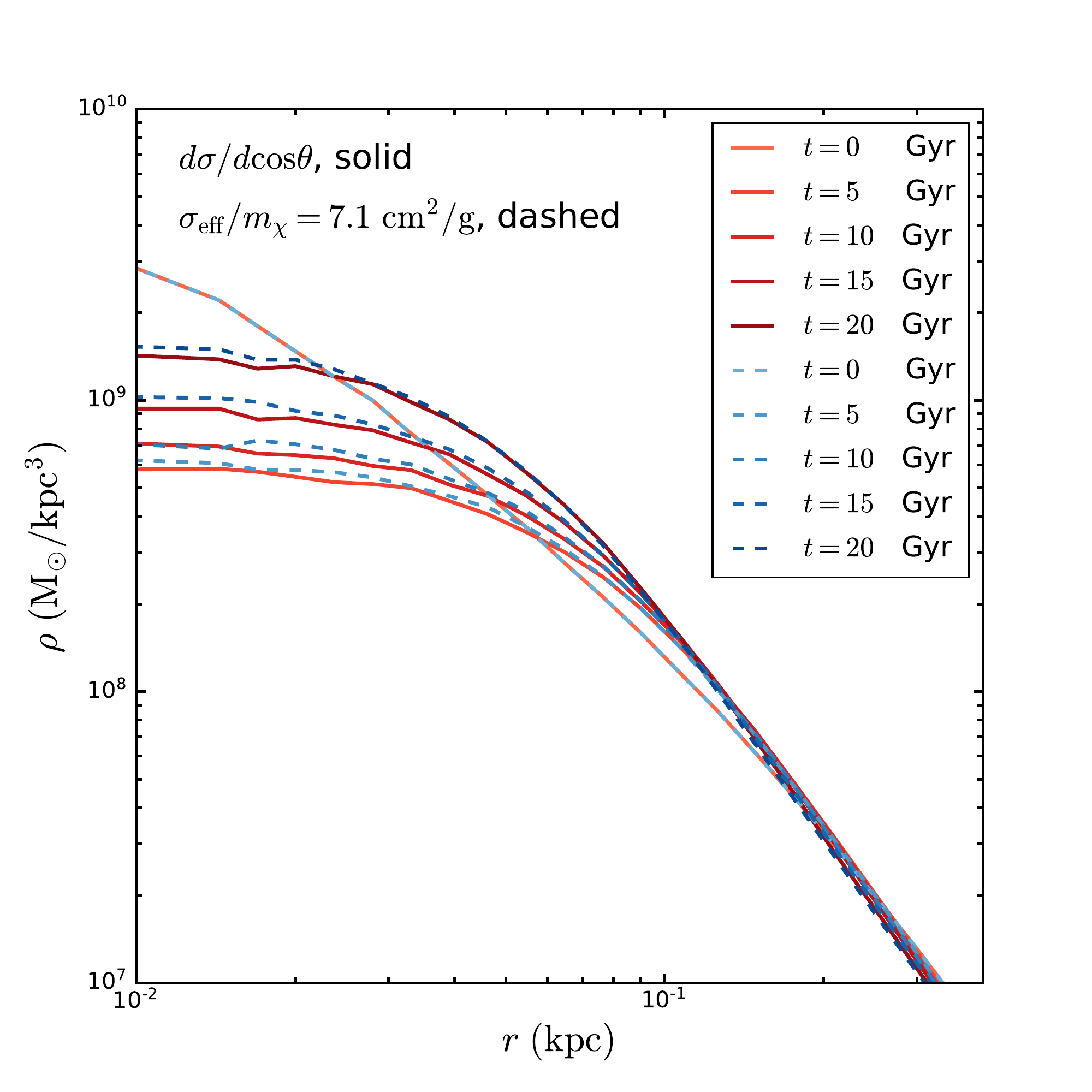}
  \includegraphics[width=7.2cm]{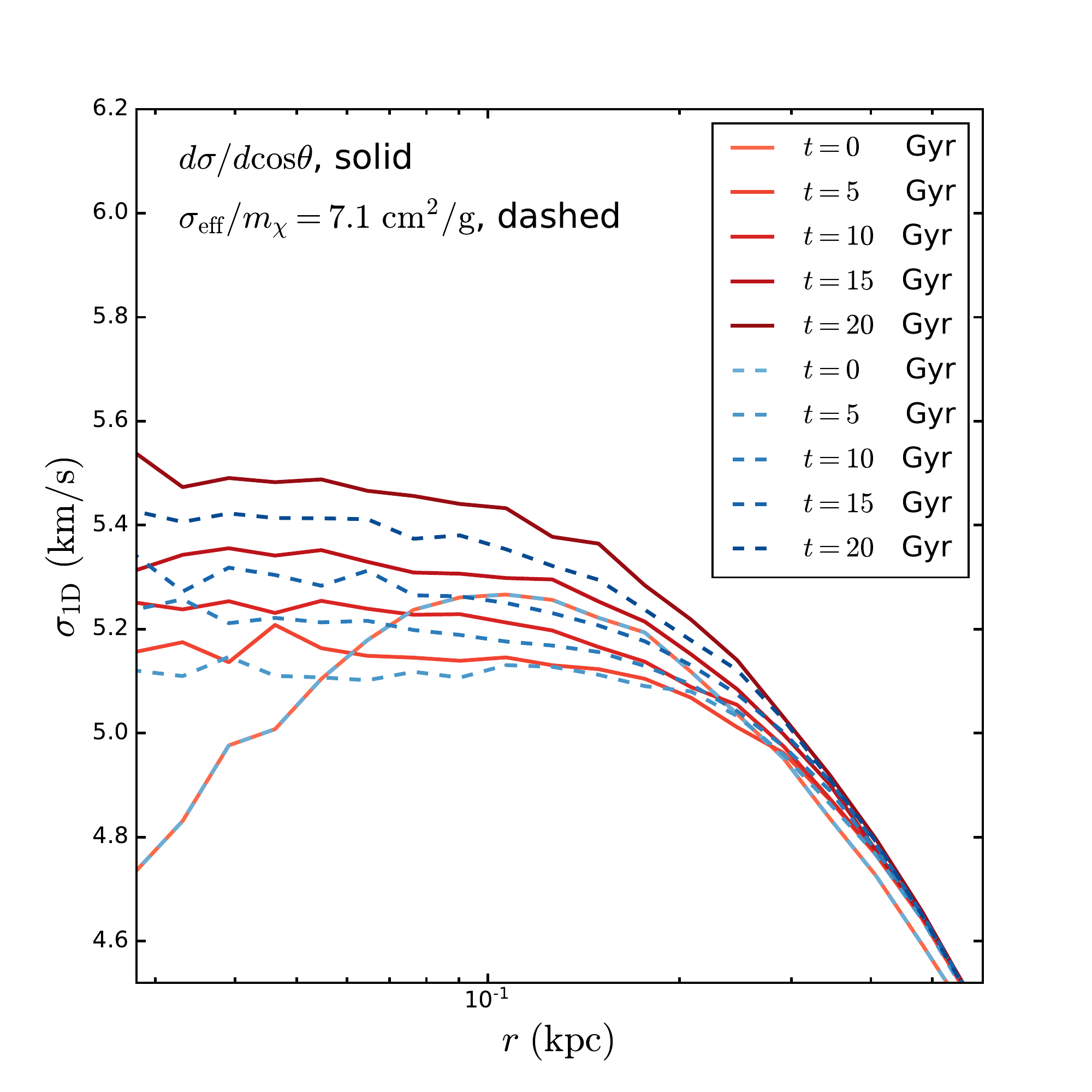}
  \caption{
    Profiles of the density (left panel) and velocity dispersion (right panel) at different evolution times from the $d\sigma/d\cos\theta$ (solid) and $\sigma_{\rm eff}/m_\chi=7.1~{\rm cm^2/g}$ (dashed) simulations. 
     \label{fig:Vprofs}
 }
\end{figure}

From Fig.~\ref{fig:sigmaeffective}, we have seen that the constant effective cross section well captures the evolution of the SIDM halo with angular- and velocity-dependent dark matter self-interactions at least up to $t\sim 20~{\rm Gyr}$. This is highly non-trivial, as with our choice of the parameters, the scatterings are mostly in the classical regime and they are largely anisotropic and strongly velocity-dependent. For the initial halo, we have deliberately chosen a high concentration, about four times the standard deviation from the cosmological median, see Table~\ref{tab:simbm}. In a realistic cosmological setup, most dark matter halos would have lower concentrations and the evolution time is the age of the universe $t\sim13.8~{\rm Gyr}$, and thus the effective cross section should provide a good approximation. 
In Fig.~\ref{fig:Vprofs}, we further show detailed profiles of the density (left panel) and velocity dispersion (right panel) from the $d\sigma/d\cos\theta$ (solid) and $\sigma_{\rm eff}$ (dashed) simulations at different evolution times. Up to $t\sim 20~{\rm Gyr}$, both simulated halos have a similar evolution history of the density and velocity dispersion profiles. 

It is interesting to test SIDM predictions for some extremely compact halos at late times in their gravothermal evolution. The central velocity dispersion further increases and $d\sigma/d\cos\theta$ decreases accordingly. Thus $\sigma_\kappa$ can be much smaller than the effective cross section calculated using $\sigma^{\rm eff}_{\rm 1D}=0.64 V_{\rm max}$. In this case, we may see differences in the halo evolution between $d\sigma/d\cos\theta$ and $\sigma_{\rm eff}/m_\chi$ simulations. 
 This is relevant for testing SIDM in extreme limits, in particular if the central halo enters the short-mean-path regime. For example, Refs.~\cite{Pollack:2014rja,Feng:2020kxv} studied a mechanism that the central region of an SIDM halo collapses into a seed black hole~\cite{Balberg:2001qg}, which could further grow into a supermassive black hole in the early universe, and they assumed a constant cross section. If the actual timescale for collapsing into the seed could be shorter than that estimated in~\cite{Pollack:2014rja,Feng:2020kxv}, and the mechanism would be further favored in explaining the origin of supermassive black holes at high redshifts. N-body simulations in the deep collapse phase are computationally expensive and we will leave the study for future work.

\subsection{Roles of dark matter self-scattering in different regions of the halo}

\begin{figure}[t]
  \centering
  \includegraphics[width=7.2cm]{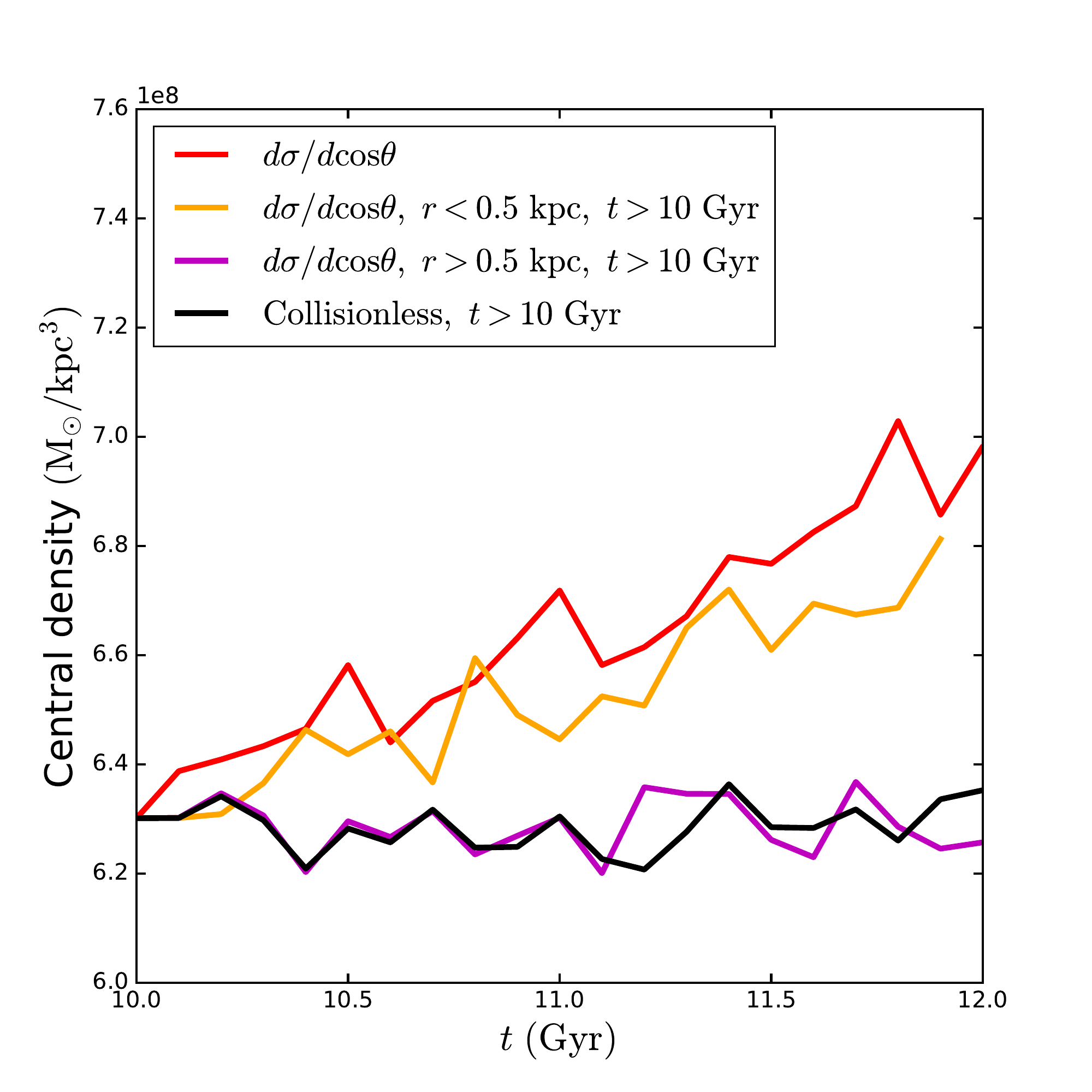}
  \caption{ 
The evolution of the central density from the $d\sigma/d\cos\theta$ simulations with dark matter self-interactions turned on only in the inner regions $r<0.5~{\rm kpc}$ (orange) or in the outer regions $r>0.5~{\rm kpc}$ (magenta), compared to the regular case where the scatterings occur in the whole region of the halo (red) and the collisionless limit (black).
  \label{fig:checkr}  
  }
\end{figure}

In Fig.~\ref{fig:sigmak} (left panel), we see that $\sigma_\kappa/m_\chi$ can be much higher than $\sigma_{\rm eff}/m_{\chi}=7.1~{\rm cm^2/g}$ in the outer regions for  $r>0.5~{\rm kpc}$. It is interesting to check how the collisions in the inner and outer regions affect the gravothermal evolution of the halo. We take the snapshot from the the $d\sigma/d\cos\theta$ simulation at $t=10~{\rm Gyr}$, and re-simulate it to $t=12~{\rm Gyr}$, while restricting the dark matter scatterings within or outside a radius of $0.5~{\rm kpc}$. As shown in Fig.~\ref{fig:checkr}, the density evolution of the simulation with the scatterings confined within $r=0.5~{\rm kpc}$ (orange) is similar to the regular case (red), although the former has slightly lower densities overall. On the other hand, when we keep the scatterings only in the outer regions $r>0.5~{\rm kpc}$ (magenta), the halo evolution is almost identical to the collisionless limit (black). Thus the gravothermal evolution of an SIDM halo is mainly driven by the scatterings in the inner regions. This justifies the use of the constant effective cross section, although it underestimates heat conductivity in the outer regions.

\subsection{Different halo initial conditions}
\label{sec:differenthalo}

\begin{figure*}
  \centering
  \includegraphics[width=4.8cm]{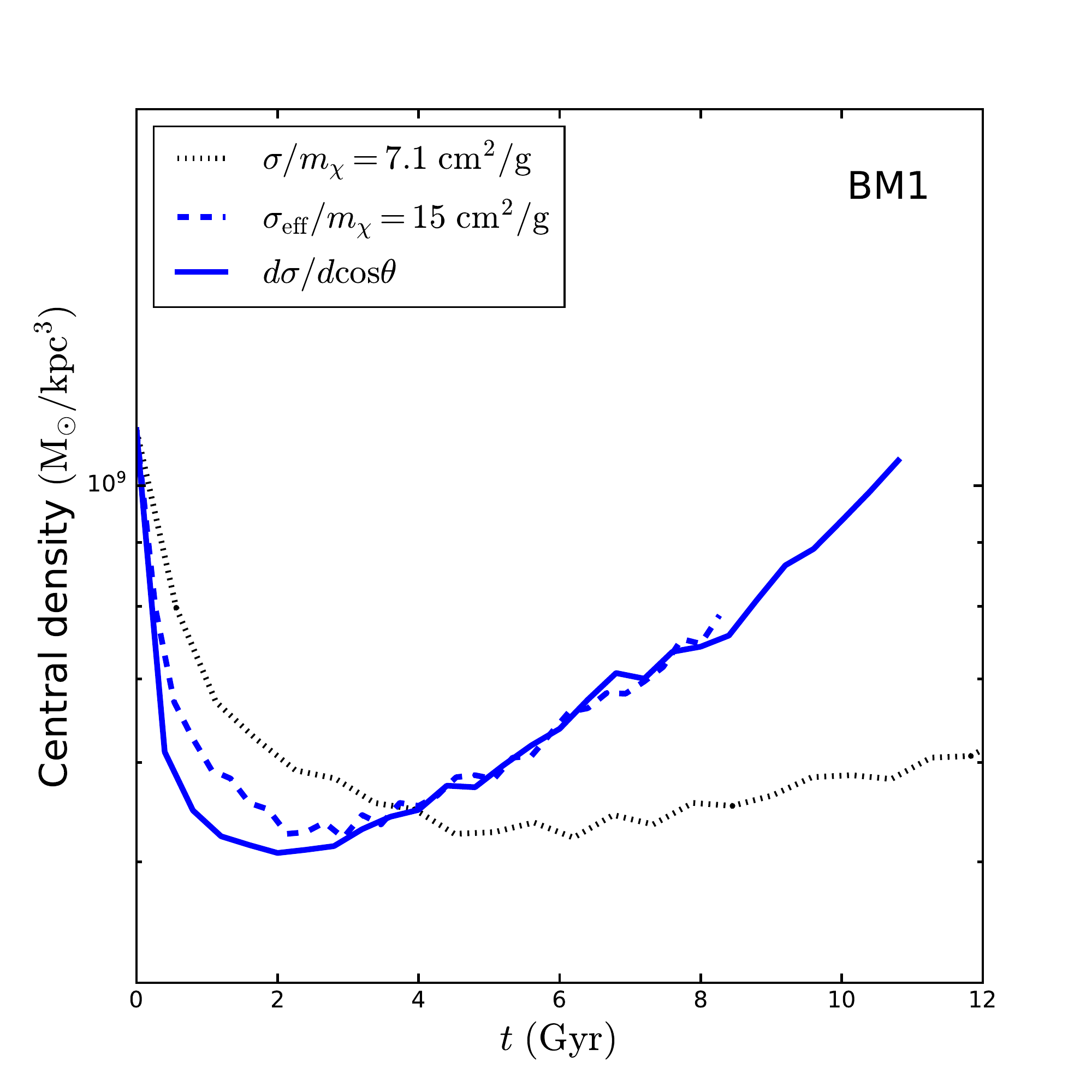}
  \includegraphics[width=4.8cm]{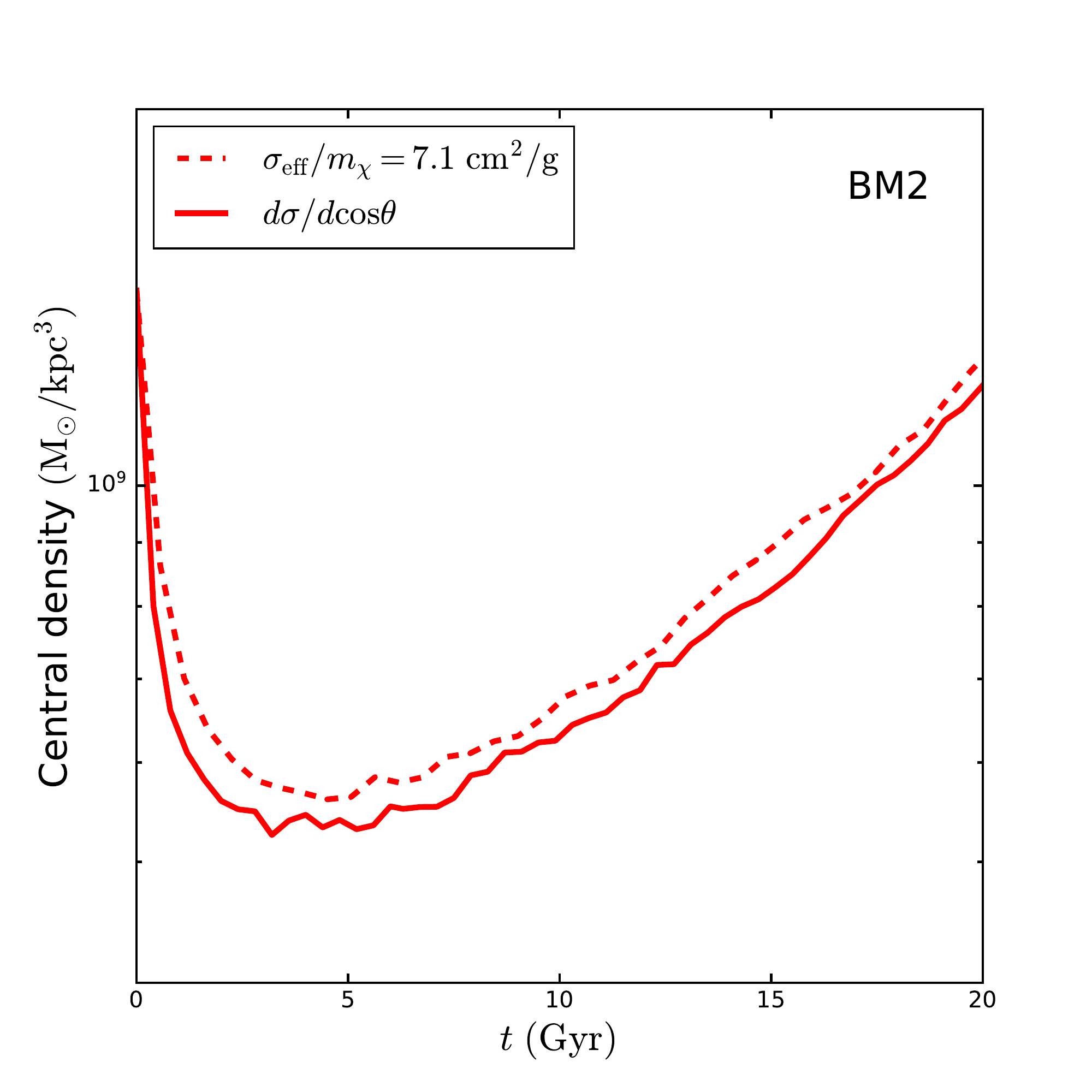}
  \includegraphics[width=4.8cm]{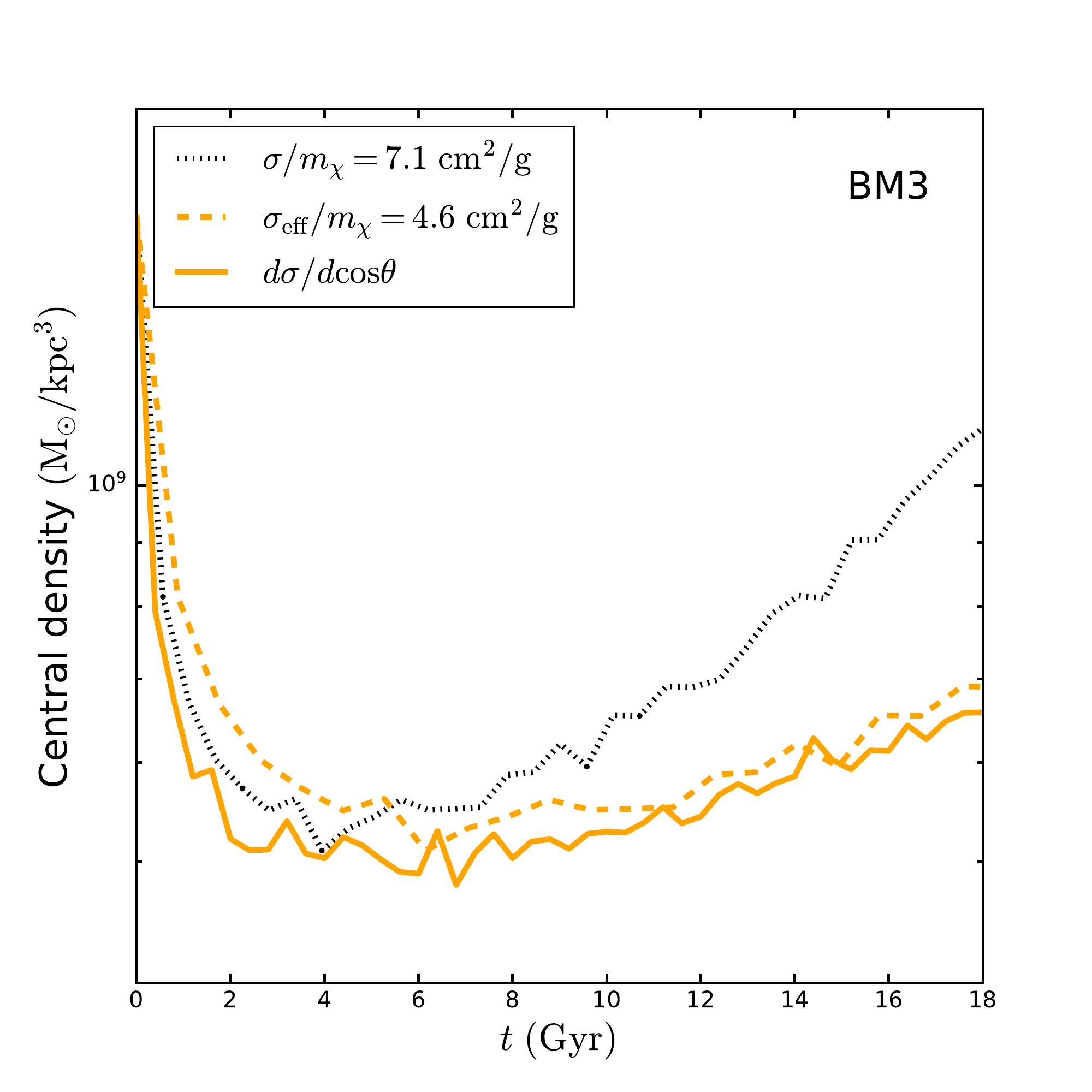}
  \caption{Gravothermal evolution of the central density from simulations with the differential cross section (solid) and the effective cross section (dashed) for the BM1 (left panel), BM2 (middle panel) and BM3 (right panel) initial conditions as listed in Table~\ref{tab:simbm}. For comparison, results with a constant cross section of $\sigma/m_\chi=7.1~{\rm cm^2/g}$ are also shown for BM1 and BM3 (dotted). 
  \label{fig:match1}
  }
\end{figure*}

The simulation results we have shown so far are based on the BM2 halos, see Table~\ref{tab:simbm}. We extend our study to two other halos. For the BM1, BM2 and BM3 halos, their 1D velocity dispersions, calculated using $\sigma^{\rm eff}_{\rm 1D}=0.64 V_{\rm max}$, are $\sigma^{\rm eff}_{\rm 1D}\approx4.1$, $5.1$ and $5.8~{\rm km/s}$, and the corresponding effective cross sections are $\sigma_{\rm eff}/m_\chi=15$, $7.1$ and $4.6~{\rm cm^2/g}$, respectively. We have taken $\sigma_0/m_\chi=2.4\times10^4~{\rm cm^2/g}$ and $w=1~{\rm km/s}$ as before.

Fig.~\ref{fig:match1} shows the evolution of the central densities for BM1 (left panel), BM2 (middle panel) and BM3 (right panel) from simulations using the differential cross section $d\sigma/d\cos\theta$ (solid) and the effective cross section (dashed). In the core-expansion phase $t\lesssim4~{\rm Gyr}$, the three halos evolve in a similar way. However, after they deeply enter the collapse phase, their central densities differ significantly, especially for $t\gtrsim8~{\rm Gyr}$. Among the three, the central density of BM1 increases fastest, while BM3 slowest. Since their masses are similar and concentrations are almost identical, the difference is mainly caused by the velocity dependence of the cross section as $d\sigma/d\cos\theta\propto v^{-4}$. BM1 has the highest effective cross section, while BM3 has the lowest. 

For the BM1 and BM3 halos, the agreement between $d\sigma/d\cos\theta$ and $\sigma_{\rm eff}$ simulations is as good as the one for BM2. For comparison, we also show the results using $\sigma/m_\chi=7.1~{\rm cm^2/g}$ for BM1 and BM3 in Fig.~\ref{fig:match1} (dotted). We see that it significantly underestimates the evolution for BM1, while overestimates for BM3, because of the mismatch with their corresponding $\sigma_{\rm eff}/m_\chi$ values. This further validates our approach with the effective cross section.  

\section{Conclusions}
\label{sec:con}

We have studied gravothermal evolution of dark matter halos with differential self-scattering. With the choice of the model parameters, the scattering is highly anisotropic and strongly velocity-dependent. We designed an SIDM module and performed a number of N-body simulations to study the evolution of an isolated halo with the differential, transfer and viscosity cross sections. Our simulations show that the viscosity cross section, which is angular-independent, provides a good approximation in modeling differential dark matter collisions for Rutherford and \Moller scatterings. This result holds in both core-expansion and -collapse phases. 

We investigated the thermodynamic properties of the simulated halo and explored its evolution history from the perspective of thermodynamics. To a good approximation, the halo is in pressure equilibrium at which gravity well balances buoyancy. We further explicitly verified the second moment of the Boltzmann equation that describes the heat transport in the halo, and showed that dark matter self-scattering is in the long-mean-free-path regime. We proposed an effective cross section, which is specified by a characteristic velocity dispersion for a given halo. Our simulations show that the effective cross section, which is velocity- and angular-independent, works well in modeling the halo evolution. 

It would be interesting to extend our work to different scenarios. For example, the presence of baryons can deepen the potential and speed up the onset of the collapse~\cite{Elbert:2016dbb,Sameie:2018chj,Feng:2020kxv}, and we could examine the validity of the viscosity and effective cross sections in simulations with baryonic potential. In addition, we could also test them in substructures, where there is a dynamical interplay between dark matter self-interactions and tidal interactions. Furthermore, we could study whether the notion of an effective cross section is valid in cosmological simulations for both main halos and subhalos. This is particularly useful for testing and constraining SIDM with observations from galaxies over a wide range of mass scales. We will leave these investigations for future work. 

Note added: During the completion of this work, a related study~\cite{Outmezguine:2022bhq} appeared, which is based on the conducting fluid model. Ref.~\cite{Yang:2022zkd} introduced an averaged cross section with the $v^5$ weighting kernel, similar to the effective cross section proposed in this work. 

\acknowledgments
We would like to thank Andrew Robertson and Yi-Ming Zhong for useful discussions. This work was supported by the John Templeton Foundation under Grant ID \#61884 (DY, HBY), the U.S. Department of Energy under Grant No. de-sc0008541 (HBY), and the Munich Institute for Astro- and Particle Physics (MIAPP) which is funded by the Deutsche Forschungsgemeinschaft (DFG, German Research Foundation) under Germany's Excellence Strategy -EXC-2094-390783311 (HBY). The opinions expressed in this publication are those of the authors and do not necessarily reflect the views of the John Templeton Foundation.

\appendix

\section{Validation of the SIDM module}
\label{app:sidm}

\begin{figure}[h]
  \centering
  \includegraphics[height=8.2cm]{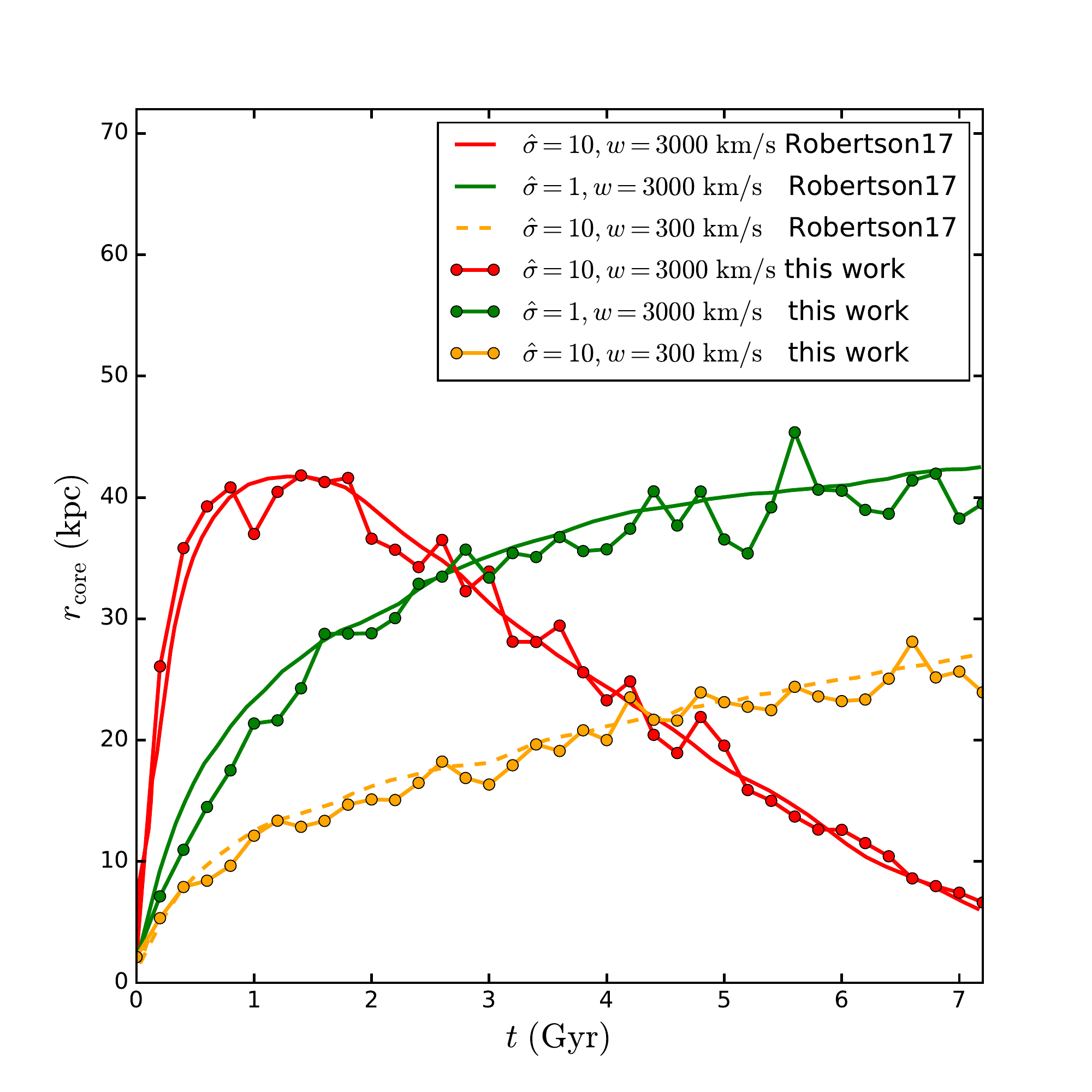}
  \caption{\label{fig:cmop3} The evolution of the core size simulated using our code, compared to the one from Ref.~\cite{Robertson:2016qef}.}
\end{figure}

To validate our SIDM module, we simulate some of the examples in Ref.~\cite{Robertson:2016qef} and show in Fig.~\ref{fig:cmop3} that they are in good agreement. 

\section{Convergence tests}
\label{app:conv}

We perform convergence tests for the SIDM simulations based on the BM2 halo. The total number of simulation particles is taken to be $N=0.1\times10^6$, $10^6$,  $4\times10^6$ and $10\times10^6$. We also investigate the role of the timestep. In the~\texttt{GADGET-2} program~\cite{Springel:2005mi}, which we use in our work, the timestep is computed as
\begin{equation}
\Delta t = \sqrt{ \frac{2\eta\epsilon}{|\mathbf{a}|} }
\label{eq:timestep}
\end{equation}
where $|\mathbf{a}|$ is the absolute value of a particle's acceleration, $\epsilon$ is the gravitational softening length, and $\eta$ is an accuracy parameter. We use the timestep defined in Eq.~\ref{eq:timestep} to calculate gravity and self-scattering probabilities, see Eq.~\ref{eq:probability}, and set the softening length to be $\epsilon=4r_{200}/\sqrt{N}$.

\begin{figure}[t]
  \centering
  \includegraphics[width=7.2cm]{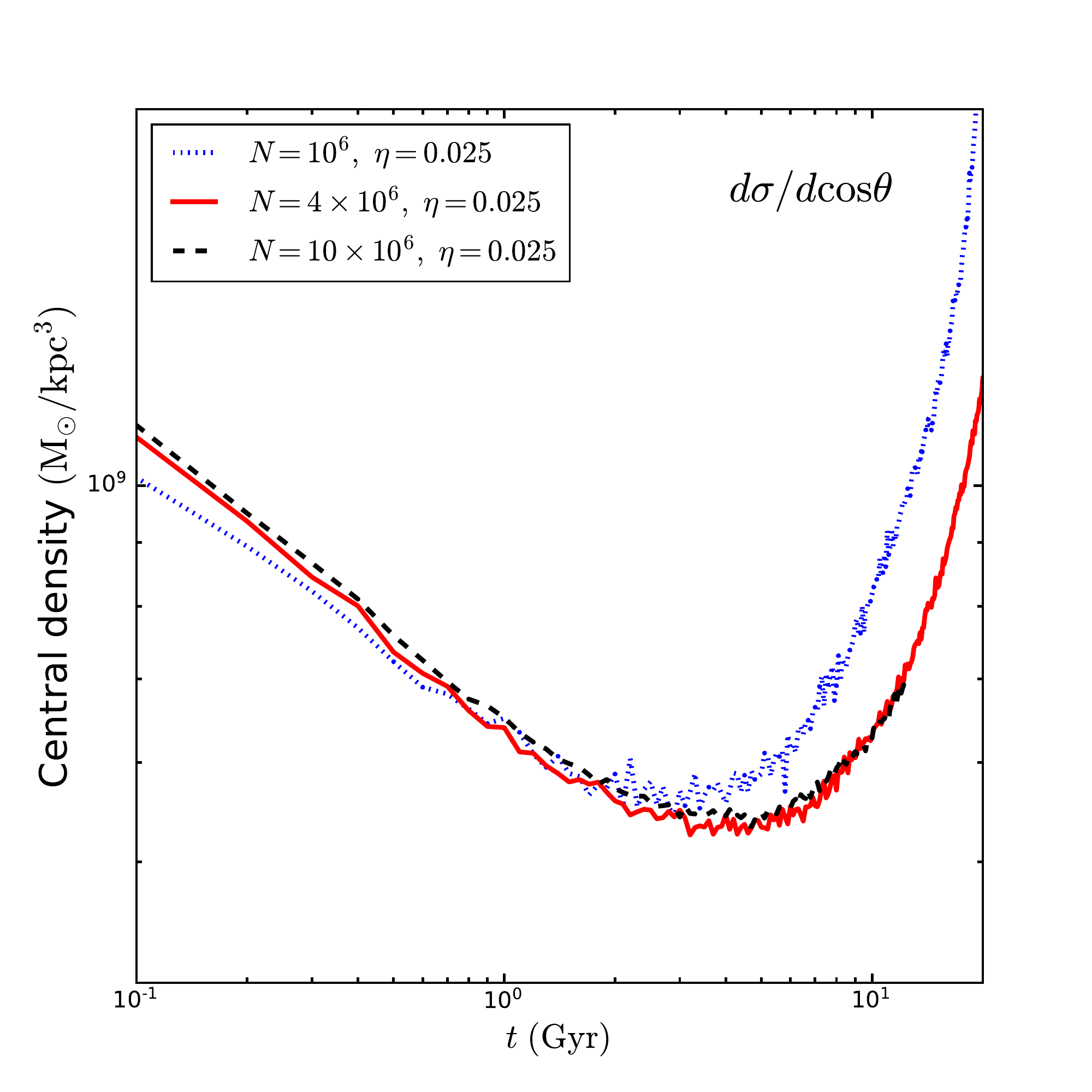}
  \includegraphics[width=7.2cm]{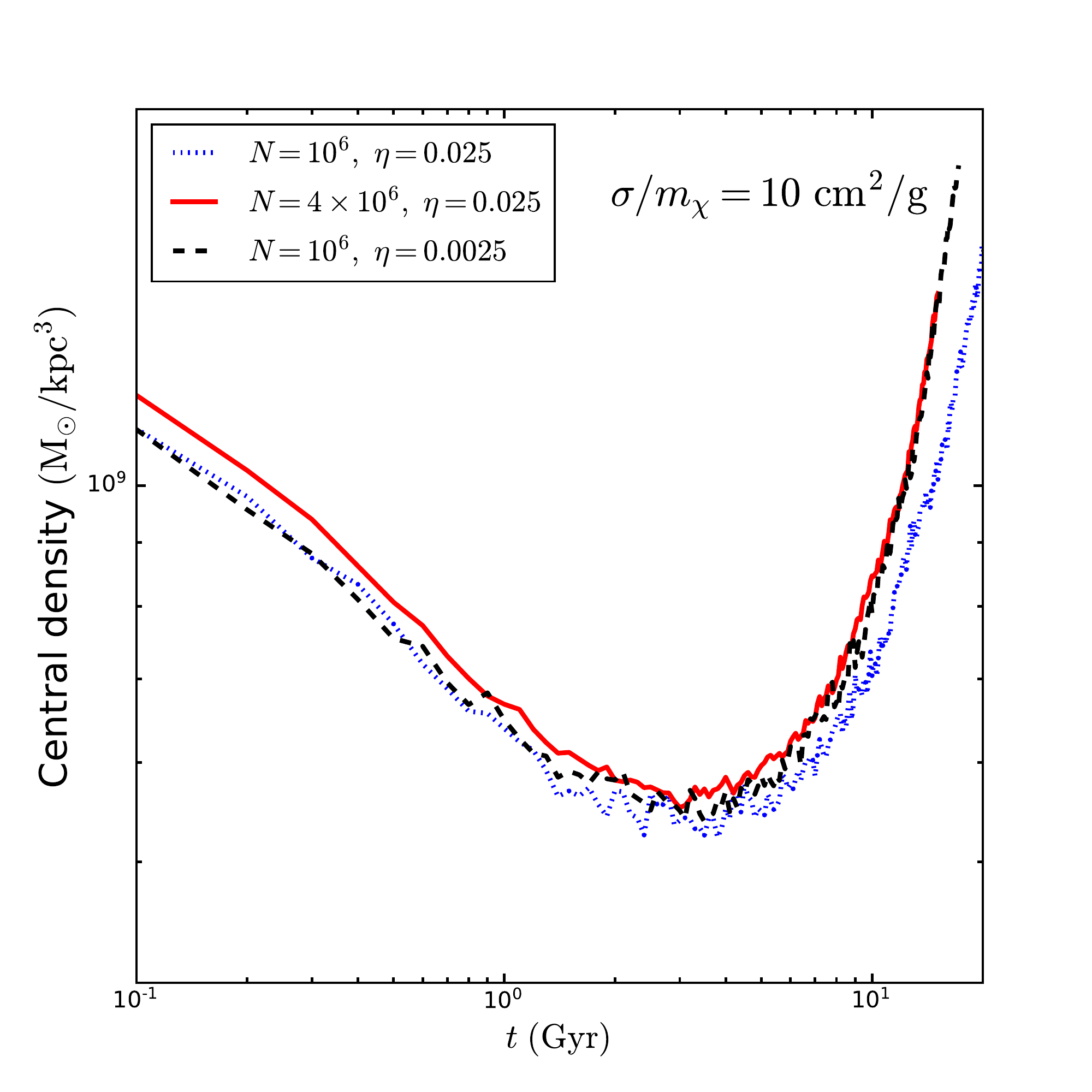}
  \caption{\label{fig:conv1} Convergence tests for $d\sigma/d\cos\theta$ (left panel) and $\sigma/m_{\chi}=10~{\rm cm^2/g}$ simulations (right panel).}
\end{figure}

Fig.~\ref{fig:conv1} (left panel) shows the evolution of the central density for $d\sigma/d\cos\theta$ simulations with $N=10^6$ (dotted blue), $4\times10^6$ (solid red) and $10\times10^6$ (dashed black), where we fix the accuracy parameter as $\eta=0.025$. We see that $N=4\times10^6$ is needed to reach the convergence for $\eta=0.025$. For $N=10^6$, the agreement in the central density is within $20\%$ for $t<10~{\rm Gyr}$, but the deviation becomes more significant for $t>10~{\rm Gyr}$ as the halo enters into the deep collapse regime. For the simulation with $N=0.1\times10^6$, the convergence is worse, and we do not report the result here. 

Fig.~\ref{fig:conv1} (right panel) shows the evolution of the central density for the $\sigma/m_\chi=10~{\rm cm^2/g}$ simulations with ($N=10^6$, $\eta=0.025$) (dotted blue), ($N=4\times10^6$, $\eta=0.025$) (solid red), as well as ($N=10^6$, $\eta=0.0025$) (dashed black). Compared to the $d\sigma/d\cos\theta$, the simulations of the constant cross section converge better in the collapse regime. The density evolution from the simulation with ($N=4\times10^6$, $\eta=0.025$) agrees well with that from the one with ($N=10^6$, $\eta=0.0025$) in particular for $t>1~{\rm Gyr}$. Given the test results, we conclude that simulations with ($N=4\times10^6$, $\eta=0.025$) or ($N=10^6$, $\eta=0.0025$) converge well for the purpose of this work. The results shown in the main text are based on high-resolution simulations that pass the convergence tests. Our results focus on a particular halo example, and it would be interesting to perform convergence tests for SIDM simulations systematically; see~\cite{Meskhidze:2022hwm} for the tests in different simulation programs. 

\begin{figure}[h]
  \centering
  \includegraphics[width=8.2cm]{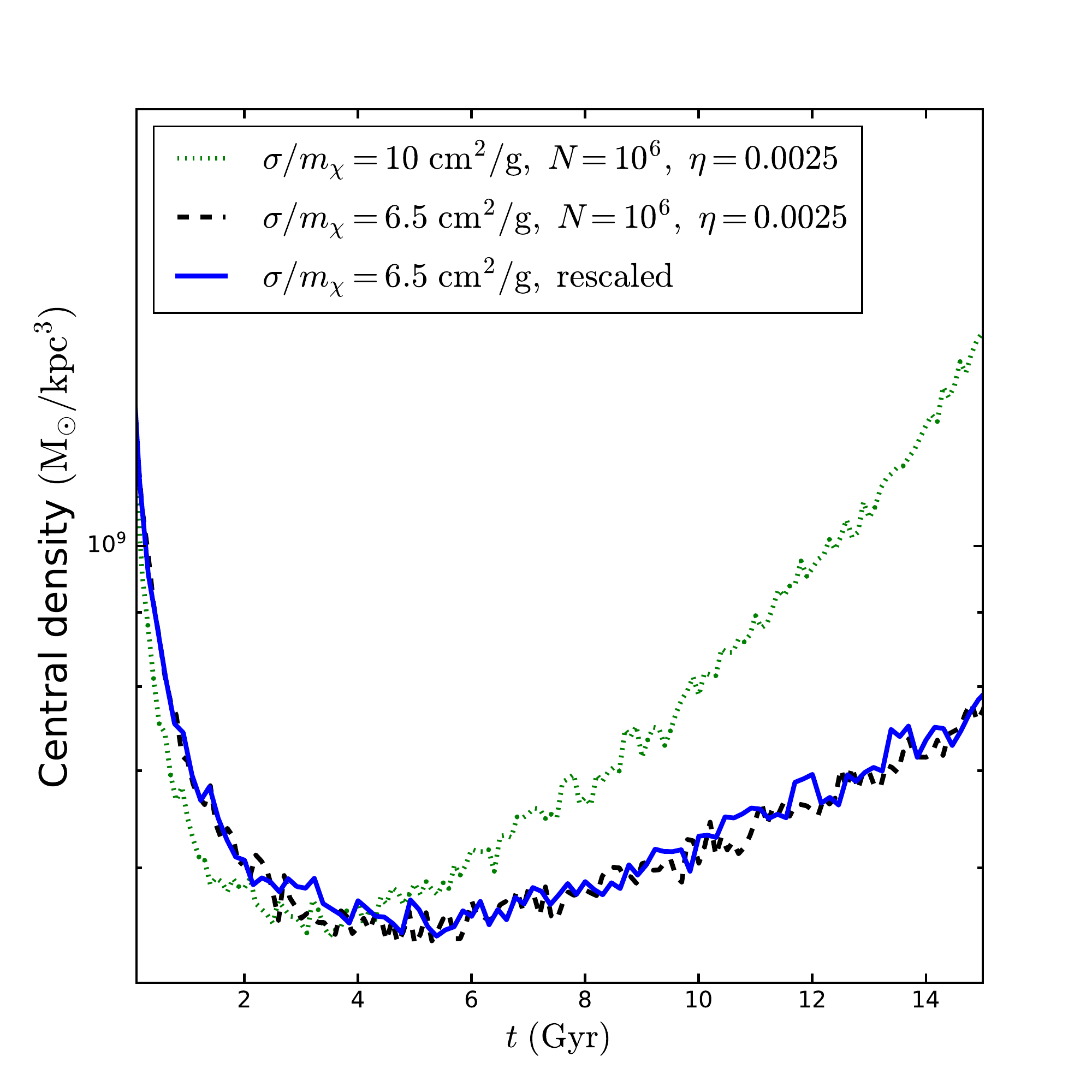}
  \caption{\label{fig:scaling} Evolution of the central density simulated for $\sigma/m_\chi=6.5~{\rm cm^2/g}$ (dashed black), compared to the one (solid blue) obtained using the $\sigma/m_\chi=10~{\rm cm^2/g}$ simulation (dotted green) after shifting $t$ by factor of $10/6.5\approx1.54$. 
}
\end{figure}

In addition, we test a rescaling method for a constant cross section. Fig.~\ref{fig:scaling} shows that the evolution of the central density simulated for $\sigma/m_\chi=6.5~{\rm cm^2/g}$ (dashed black), compared to the one (solid blue) obtained using the the $\sigma/m_\chi=10~{\rm cm^2/g}$ simulation (dotted green) after shifting $t$ by factor of $10/6.5\approx1.54$. In order to save computational time, we will use this method to obtain results for a constant cross section different from $\sigma/m_\chi=10~{\rm cm^2/g}$.

\bibliographystyle{apsrev}
\bibliography{reference}

\end{document}